\documentstyle[11pt]{article}
\def \ie {{\it i.e.}}
\def \eg {{\it e.g.}}
\def \D {\mbox{D}}

\def\be {\begin{equation}}
\def\ee {\end{equation}}

\def\la {\langle}
\def\ra {\rangle}
\def\p {\partial}
\def\bi {\bibitem}

\def\case#1/#2{\textstyle\frac{#1}{#2} }
\begin{document}

\title{1+3 Covariant Cosmic Microwave Background anisotropies I: \\
Algebraic relations for mode and multipole representations.}

\author{Tim Gebbie  \dag\ and George F.R. Ellis \dag\ \ddag\ \\ 
\\ \dag\ {\it Department of Mathematics and Applied Mathematics}, \\
  {\it University of Cape Town, Rondebosch 7701, South Africa} \\
  and \\ \ddag\ 
  {\it Department of Mathematics, Queen Mary and Westfield College}, \\
  {\it Mile End Road, London E1 4NS, UK}}

\maketitle

\begin{abstract}

This is the first of a series of papers extending a 1+3 covariant and  
and gauge invariant treatment of kinetic theory in curved
space-times to a treatment of Cosmic Background Radiation (CBR) temperature
anisotropies arising from inhomogeneities in the early universe. This paper
deals with algebraic issues, both generically and in the context of models
linearised about Robertson-Walker geometries. 

The approach represents radiation anisotropies by Projected
Symmetric and Trace-Free tensors. The Angular correlation functions for the
mode coefficients are found in terms of these quantities, following the
Wilson-Silk approach, but derived and dealt with in 1+3 covariant and
gauge invariant (CGI) form. The covariant
multipole and mode-expanded angular correlation functions are related to the
usual treatments in the literature. The CGI mode expansion is related to the
coordinate approach by linking the Legendre functions to the Projected
Symmetric Trace-free representation, using a covariant addition theorem for
the tensors to generate the Legendre Polynomial recursion relation. 

This paper lays the foundation for further papers in the
series, which use this formalism in a CGI approach to developing solutions of
the Boltzmann and Liouville equations for the CBR before and after
decoupling, thus providing a unified CGI derivation of the variety of
approaches to CBR anisotropies in the current literature.

\end{abstract}
\tableofcontents


\section{Introduction}


Ellis, Treciokas and Matravers (ETM) \cite{ETMa,ETMb} introduced a $1+3$ 
covariant kinetic theory formalism in which an irreducible representation of
the rotation group based on Projected Symmetric and Trace-Free (PSTF) tensors 
orthogonal to a physically definved 4-vrelocity $u^a$ 
gives a covariant representation of the Cosmic Background Radiation (CBR)
anisotropies, which is gauge-invariant when the geometry is an
almost-Robertson Walker (RW) geometry. This $1+3$ Covariant and Gauge-Invariant 
(CGI) formalism has been used in a previous series of papers
\cite{SME,MSEa,MSEb,SAG} to look at the local generation of CBR
anisotropies by matter and spacetime inhomogeneities and anisotropies
in an almost-Friedmann Lema\^{i}tre (FL) universe model \footnote{
Here `Robertson-Walker' refers to the geometry, whatever the field
equations; `Friedman-Lemaitre'\ assumes that the Einstein gravitational
field equations with a perfect fluid matter source are imposed on such a
geometry.}. By contrast, the present series of papers\footnote{And a series by
Challinor and Lasenby \cite{cl}, \cite{cl2}
which are similar in method and intent but
different in detail and focus.} uses this formalism to
investigate CBR anisotropies in the non-local context of emission of
radiation near the surface of last scattering in the early universe and its
reception here and now (the Sachs-Wolfe (SW) effect and its further 
developments).

There is of course a vast literature investigating these anisotropies both
from a photon viewpoint, developing further the methods of the original
Sachs-Wolfe paper \cite{SW}, and from a kinetic theory viewpoint, so it
is useful to comment on why the CGI philosophy and programme \cite{EB} 
make the present series of papers 
worthwhile. Rather than beginning with a background described in 
particular coordinates and perturbing away from this
background, this approach centres on $1+3$ covariantly defined geometric
quantities, and develops exact nonlinear equations for their evolution.
These equations are then systematically linearised about a
Friedmann-Lemaitre (FL) background universe with a Robertson-Walker (RW)
geometry resulting in description by gauge-invariant variables and equations 
\cite{EB}. Because the definitions and equations used are
coordinate-independent, one can adopt any suitable coordinate or tetrad
system to specialise the tensor equations to specific circumstances when
carrying out detailed calculations; a harmonic or mode analysis can be
carried out at that stage, if desired.

This approach is geometrically transparent (see \cite{cl}, 
\cite{cl2}) because of the CGI variable definitions used. 
In contrast to the various gauge-dependent approaches to 
perturbations in cosmology, the differential equations used are of just the 
order that is needed to describe the true physical degrees of freedom, so
no non-physical gauge modes occur. When a harmonic decomposition is 
introduced in the case of linear perturbations, the CGI variables used here
provide a description that is equivalent to that obtained by approaches 
based Bardeen's GI variables \cite{bardeen}, see \cite{bde}, but they do 
not imply linearisation of the equations from the outset, as occurs in 
that formalism.

Thus the benefit of the present formalism is precisely its $1+3$ covariant and
gauge invariant nature, together with the fact that we are able to write
down the exact non-linear equations governing the growth of structure and
the propagation of the radiation, and then linearise them in a transparent
way in an almost-RW situation. This means it can be extended to non-linear
analyses in a straightforward way \cite{MGE}, which will be essential in 
developing the
theory of finer CBR anisotropy structure as reliable small-angle
observations become available. Achievements of the CGI approach with respect 
to the CMB are the
almost-EGS Theorem \cite{SME}, related model-independent limits on 
inhomogeneity and isotropy \cite{MSEa,MSEb,MES96,SAG}, and derivation 
of exact anisotropic solutions of the Liouville equation in a RW 
geometry (\cite{me}, see also \cite{emt}).

This paper, Part I, deals with algebraic issues, developing further the
formalism of ETM: namely an irreducible representation of radiation
anisotropies based on PSTF tensors \cite{T1,P}. The paper considers 
this irreducible representation and its relation to observable quantities, 
both generically and in the context of models linearised about RW 
geometries \cite{MSEa}. In section 2 and 3, the underlying $1+3$ 
decomposition is outlined and the basic CGI harmonic formalism for 
anisotropies developed. In section 4, the angular correlation
functions are constructed from CGI variables, assuming that the multipole
coefficients are generated by superpositions of homogeneous and isotropic
Gaussian random fields. The multipole expansion is discussed in detail,
extending the results of ETM, giving the construction of the multipole
coefficient mean-square and developing its link to the angular correlation
function. In section 5, the mode coefficients are found following the
Wilson-Silk approach, but derived and dealt with in the CGI form; the
covariant and gauge invariant multipole and mode expanded angular 
correlation functions are related to the usual treatments used in the 
literature \cite{WS,GSS,HS95a,HS95b,EBW}. In this discussion, the
CGI mode expansion is related to the
coordinate approach by linking the Legendre Tensors to the PSTF 
representation, using a covariant addition theorem 
to generate the Legendre Polynomial recursion relation. The key result 
is the construction of the angular correlation functions in the 
CGI variables, and their link to the 
(non-local) GI Mode functions \cite{GSS}.

The following papers in the series look at the Boltzmann equation and
multipole divergence relations, solution of the resulting mode equations,
and relation of the kinetic theory approach to the photon based formalism of
the original Sachs-Wolfe paper. Exact non-linear equations are obtained and
then linearised, allowing a transparent linearisation process from the
non-linear equations that is free from ambiguities and gauge modes.


\section{Temperature anisotropies}

A radiation {\it temperature measurement} is associated with an antenna
temperature, $T(x^i,e^a)$, measured by an observer moving with
4-velocity $u^a$ at position $x^i$ in a direction $e^a$ on the unit
sphere ($e^ae_a=1$, $e^au_a=0)$. We assume $u^a$can be uniquely define
in the cosmological situation, corresponding to the motion of
`fundamental observers' in cosmology \cite{varenna} \footnote{In the
early universe, when matter and radiation average velocities differ,
there may be several competing possibilities for covariant definition
of $u^a$; however once a choice has been made between these
possibilities, this vector field is uniquely defined.}. The direction
$e^a$ can be given in terms of an orthonormal tetrad
frame\footnote{The description of such a tetrad frame is briefly
discussed in last appendix}, for example by : 
\begin{equation}
e^a(\theta ,\phi )=(0,\sin \theta \sin \phi ,\sin \theta \cos \phi
,cos\theta )\,.\label{3.2}
\end{equation}
The temperature $T(x^i,e^a)$ can be unambiguously decomposed into the
all-sky {\it average bolometric temperature } \footnote{%
Note that this is {\it not} the same as the background temperature, for that
quantity varies only with cosmic time $t$, whereas the true isotropic
component of the temperature varies with spatial position as well as time.} $%
T(x^i)$ at position $x^i$, given by 
\begin{equation}
T(x^i)=\frac 1{4\pi }\int_{4\pi }T(x^i,e^a)d\Omega .  \label{3.6}
\end{equation}
where $\Omega $ is the solid angle on the sky, and the anisotropic {\it %
temperature perturbation} $\delta T(x^i,e^a)$ (the difference from the
average over the unit sphere surrounding $x^i$ \cite{MSEa}),
can be defined : 
\begin{equation}
T(x^i,e^a)=T(x^i)+\delta T(x^i,e^a). 
\label{3.5}
\end{equation}
{}From the Stefan-Boltzmann law it follows -- if the radiation is almost
black-body, which we assume -- that the radiation energy density is given in
terms of the average bolometric temperature by $\rho_R
(x^i)=r T^4(x^i)$.
($r$ is Stefan-Boltzmann constant). Both the quantities $T(x^i)$ and $\delta
T(x^i,e^a)$ are CGI, for $T(x^i)$ is defined
in a physically unique frame in the real universe 
(because $u^a$ is assumed to be uniquely defined), 
and $\delta T(x^i,e^a)$ vanishes in any background without temperature
anisotropies.

We can define the fractional temperature variation $\tau (x^i,e^a)$ by
\cite{MSEa}
\begin{equation}
\tau (x^i,e^a):={\frac{\delta T(x^i,e^a)}{T(x^i)}},  \label{3.10}
\end{equation}
and take a covariant (angular) harmonic expansion of this, 
\begin{equation}
\tau (x^i,e^a)=\sum_{\ell=1}^\infty \tau
_{a_1a_2a_3...a_\ell}(x^i)\,e^{a_1}e^{a_2}e^{a_3}...e^{a_{\ell-1}}e^{a_\ell}
\equiv \sum_{\ell \geq 1} \tau_{A_{\ell}} \hat e^{A_{\ell}}.
\label{3.11}
\end{equation}
We introduce the shorthand notation using the compound index 
$A_{\ell} = a_1 a_2 ... a_{\ell}$. Here $\tau _{a_1a_2a_3...a_l}(x^i)$
are trace-free symmetric tensors orthogonal to $u^a$ :
\begin{eqnarray}
\tau_{A_{\ell}}=\tau_{(A_{\ell})}\,,~~\tau_{A_{\ell} ab} h^{ab}
=0\,,~~\tau_{A_{\ell} a} u^a \label{symmetry}
\end{eqnarray}
Round bracketts ``(..)'' denote the symmetric part of a set of
indices, angle bracketts ``$\la..\ra$'' the (orthogonally-) Projected 
Symmetric Trace-Free (PSTF) part of the indices : $\tau_{A_{\ell}} =
\tau_{\la A_{\ell} \ra}$.

Because of (\ref{3.2}), this expansion is entirely equivalent to a more
usual expansion in terms of spherical harmonics:
\begin{equation}
\tau (x^i,e^a)=\sum_{l=1}^\infty A_l^m(x^i)Y_l^m(\theta ,\phi )
\label{usual}
\end{equation}
(see \cite{ETMa} for details), but is more closely related to a tensor
description, and so results in more transparent relations to physical
quantities. 

We wish to measure the temperature in two different directions to find the
temperature difference associated with the directions $e^a$ and $e^{\prime
}{}^a$ such that (using \ref{3.6}) : 
\begin{eqnarray}
\Delta T(x^i;e^a,e^{\prime }{}^a)=T(x^i,e^a)-T(x^i,e^{\prime }{}^a)\,
~\Rightarrow~ \Delta T(x^i;e^a,e^{\prime }{}^a)
=\delta T(x^i,e^a)-\delta T(x^i,e^{\prime}{}^a).  \label{3.8}
\end{eqnarray}
It follows from (\ref{3.8}), (\ref{3.10}) and (\ref{3.11}) that 
 ${{\Delta T(x^i;e^a,e^{\prime }{}^a)}} = T(x^i) \sum_{\ell} 
\tau _{A_{\ell}}\left( { e^{A_{\ell}}-e^{\prime}{}^{A_{\ell}}}\right)$\,,
where $\Delta T/T$ represents the {\it real fractional temperature
difference} on the current sky. Due to the CGI nature
of $T(x^i)$ we may relate this directly to the {\it real temperature 
perturbations} (no background model is involved in these definitions).

The relation between the two directions $e^a$ and $e^{\prime }{}^a$ at $x^i$
is characterised by 
\begin{equation}
e^ae^{\prime }{}_a=\cos (\beta )=:X.  \label{3.1}
\end{equation}
i.e. they are an angular distance $\beta $ apart \footnote{%
It should be pointed out that $e^{\prime}{}^a$ is distinct from $%
e^{a^{\prime}{}} $, the first denotes a direction vector different from $e^a$
in a given tetrad frame, while the second means the same direction vector in
a different tetrad frame.}. If analogous to (\ref{3.2}) we write 
$e^a(\theta ^{\prime }{},\phi ^{\prime }{})\equiv {e^{\prime }{}}^a=(0,\sin
\theta ^{\prime }{}\sin \phi ^{\prime }{},\sin \theta ^{\prime }{}\cos \phi
^{\prime }{},\cos \theta ^{\prime }{})$, then it can be shown from 
(\ref{3.1}) that 
\begin{equation}
\cos \beta =\sin \theta \sin \theta ^{\prime }{}\cos (\phi -\phi ^{\prime
}{})+\cos \theta \cos \theta ^{\prime }{}.
\end{equation}
In later applications, it is important to relate the different terms of the
harmonic expansion to angular scales in ths sky. A useful approximation is
$l\approx {\frac 1\theta }$, where $\theta $ is in radians. 

\subsection{Covariant and gauge invariant angular correlation function}

The two-point correlations are an indication of the fraction of temperature
measurements, $T(x^i,e^a)$, that are the same for a given angular
separation. This corresponds to the correlation between $\delta T(x^i,e^a)$
and $\delta T(x^i,e^{\prime }{}^a)$ or equivalently between $\tau (x^i,e^a)$
and $\tau (x^i,e^{\prime }{}^a)$, given by the {\it angular position
correlation function} 
\begin{equation}
C({e^a,e^{\prime }{}^a})=\left\langle {\tau (x^i,e^a),\tau (x^i,e^{\prime
}{}^a)}\right\rangle ,  \label{ang_correl}
\end{equation}
where the angular brackets representing an angular average over the complete
sky. Note this is a function in the sky.{} If we write ${\tau (x^i,e^a)}$
and ${\tau (x^i,e^{\prime }{}^a)}$ in terms of the angular harmonic
expansion (\ref{3.11}), we can also define correlation functions $C_l$
for the anisotropy coefficients ${\tau _{A_l}(e}^a),{\tau }_{A_l}(e^{\prime
a})$ by 
\begin{eqnarray}
C_\ell({e^a,e^{\prime }{}^a})=(2\ell+1)^{-1}\Delta _\ell\left\langle {\tau
_{A_\ell}(x^i,e^a) \tau ^{A_\ell}(x^i,e}^{\prime }{^a)}\right\rangle .
\label{129a}
\end{eqnarray}
Here the right-hand side term in brackets is the all-sky mean-square value
of the $l-$th temperature coefficient ${\tau _{A_\ell}(x^i,e^a)}$, and the
coefficient $\Delta _\ell$ is defined in (\ref{Delta-ortho0-01}). The numerical
factor $(2\ell+1)^{-1}\Delta _\ell$ is included in order to agree with
definitions normally used in the literature (see later). This can be
thought of as the momentum space version of (\ref{ang_correl}), as we
have taken an angular fourier series of the quantities in that equation;
it says, for each choice of ${e^a,e^{\prime }{}^a,}$ how much power
there is in that expression for that angular separation as contributed
by a particular $\ell$-th valued multipole moment on average.

\subsection{The Central-Limit Theorem}

We consider an ensemble of temperature anisotropies, where a 
sequence of repeated trials is replaced by a
complete ensemble of outcomes. The temperature
anisotropy $\tau_{A_\ell}$ found in a given member of the ensemble is 
a realization of the statistical process represented by the ensemble.
The physically measured anisotropy is taken to be 
one such realization. The variance of the ensemble, for example
$\langle \tau_{A_\ell} \tau^{A_\ell} \rangle$, is in principle 
found by averaging over a sufficiently large number of
experiments, where we assume the results will approach the
true ensemble variance -- this is the assumption of
ergodicity. 

On Fourier transforming, we make the assumption that to a good 
approximation the phases of the various multipole moments are 
uncorrelated and random. This corresponds to treating the anisotropies 
as a form of random noise. The random phase assumption has a useful 
consequence: that the sum of a large number of independent random variables 
will tend to be normally distributed. By the central-limit theorem 
\cite{Adler81}, this is true for all quantities that are derived from 
linear sums over waves. The end result is that one ends up with a 
Gaussian Random Field (GRF) which is fully characterized by a power 
spectrum. The central limit theorem holds as long as there exists a 
finite second moment, i.e. a finite variance.

We will assume that the angular variance 
$\langle \tau_{A_\ell} \tau^{A_\ell} \rangle$ is independent of 
position; this is the assumption of statistical homogeneity. 
Its plausibility lies in the underlying use of the weak Copernican 
assumption. Additionally it is convenient to assume that the power 
spectrum will have no directional dependence, thus it will be isotropic:
$P(k^a) = P(|k^a|)$.  Together these imply the statistical distribution 
respects the symmetries of the RW background geometry.

One needs to be careful in using the central limit theory to motivate
Gaussian random field, particularly in the presence of nonlinearity, which 
could result in the elements of the ensemble no longer being 
independent. While the assumptions of primordial homogeneous and 
isotropic GRF's is plausible, because the perturbations
 are made up of a sufficiently large number
of independent random variables, the key point is to realize that 
these are assumptions that should be tested if possible.
The simplest test of weak non-Gaussianity is 
looking for a three-point angular or spatial correlation, for
the Gaussian assumption ensures that all the odd higher moments 
are zero and that the even ones can be expressed in terms of the 
variance alone. 
If the primordial perturbations are made up of GRF's, 
then non-Gaussianity of the CBR
anisotropy spectrum should arise primarily from foreground 
contamination due to local physical processes.
If the non-Gaussian effects due to these later physical processes or 
evolutionary effects are small enough, one can attempt to determine
a cosmological primordial signature. 
     
\subsection{Gaussian perturbations}

A general {\it Gaussian perturbation} \cite{LW}, $\tau (x^i,e^a)$, will
be a superposition of functions, $\tau _{A_\ell}$, i.e. (\ref{3.11}) 
is satisfied, where the probability, P, of finding a particular valued
temperature coefficient is given by ($\sigma^2_{\ell} = \la
\tau_{A_\ell} \tau^{A_{\ell}} \ra$) :
\begin{equation}
P(\tau _{A_\ell})={\frac 1{\sqrt{2\pi \sigma _\ell^2}}}\exp 
\left\{ {\frac{-\tau_{A_\ell}\tau ^{A_\ell}}{2\sigma _\ell^2}}\right\}
.  \label{4.1} 
\end{equation}
Note that{\bf \ }$\tau _{A_\ell}${\bf \ }is both the amplitude of the
$\ell$-th component, and determines the probability of that
amplitude. The probability of a temperature perturbation, $\tau $, is
given by the sum of the Gaussian probability distributions (\ref{4.1})
weighting the various angular scales, given by $\ell$, of the general
perturbation (\ref{3.11}). 

Considering isotropic and homogeneous gaussian random fields, the angular
position correlation function $C({e^a,e^{\prime }{}^a})$ is a function only
of the angular separation $\beta $ of the two temperature measurements. We
then write (\ref{ang_correl}) as 
\begin{eqnarray}
\left\langle {\tau \cdot \tau ^{\prime }{}}\right\rangle =C(\beta )=W(X)\,,
\label{w}
\end{eqnarray}
where the expression on the left is shorthand for $\left\langle {\tau
(x^i,e^a),\tau (x^i,e^{\prime }{}^a)}\right\rangle _\beta $ , the {\it %
2-point angular correlation function} for a given angular separation $\beta $
between the on-sky temperature measurements, and $X$ $=e^ae^{\prime
}{}_a=\cos \beta $. This expression is now independent of position in the
sky. Gaussian Fields are completely specified by the angular power spectrum
coefficients $C_\ell$ (\ref{129a}), which are now just constants, 
because $\ell$ is uniquely related to $\beta$, so the power spectrum
is a function of the modulus of the wavenumber only. One thus expects 
the temperature perturbations in this case to be fully specified by
the {\it mean squares}, $\left\langle \tau _{A_\ell}\tau
^{A_\ell}\right\rangle $, when (\ref{3.11}) is
substituted in (\ref{w})$.$ Equivalently they are uniquely determined by
the angular Fourier transform of the 2-point angular correlation function .


\section{Multipole expansions}


In this section we examine the anisotropy properties of radiation described
in terms of the covariant multipole formalism (\ref{3.11}), which is
equivalent to the usual angular harmonic formalism but much more directly
related to space-time tensors. Note that the relations in this section hold
at any {\it point} in the space-time, and in particular at the event $R$
('here and now') where observations take place. Here we consider the
PSTF part of, $e^{A_{\ell}}$; some useful properties of $e^{A_{\ell}}$
are listed in appendix B.

\subsection{The PSTF part of $e^{A_{\ell}}$}

Because the coefficients in (\ref{3.11}) are symmetric and trace-free,
the important directional quantities defined by directions $e^a$ at a
position $x^i$ are the PSTF quantities 
\begin{equation}
O^{A_\ell }=e^{\left\langle {A_\ell }\right\rangle
}=e^{\la a_1}e^{a_2}e^{a_3}...e^{a_{\ell -1}}e^{a_\ell \ra},  \label{defO}
\end{equation}
for clearly 
\begin{equation}
\tau (x^i,e^a)=\sum_{\ell =1}^\infty \tau _{A_\ell }(x^i)e^{A_\ell
}(\theta ,\phi 
)=\sum_{\ell =1}^\infty \tau _{A_\ell }(x^i)O^{A_\ell }(\theta ,\phi
).  \label{3.11b} 
\end{equation}
Indeed the standard spherical harmonic properties are contained in these
quantities.

Now the symmetric trace-free (STF) part of a 3-tensor is given in general by
Pirani \cite{P}, 
\begin{eqnarray}
\left[ {F_{A_\ell }}\right]
^{STF}=\sum_{n=0}^{[\ell
/2]}B^{ln}h_{(a_1a_2}....h_{a_{2n-1}a_{2n}}F_{a_{2n+1}...a_\ell )}~~ 
\mbox{with}~~
B_{\ell n}={\frac{{(-1)^n\ell !(2\ell -2n-1)!!}}{{(\ell -2n)!(2\ell
-1)!!(2n)!!}}}.  \label{B} 
\end{eqnarray}
Here $[\ell /2]$ means the largest integer part less than or equal to
$\ell /2$. The following definitions have also been used : $\ell !
=\ell (\ell -1)(\ell -2)(\ell -3)...(1)$, and $\ell !! =\ell (\ell
-2)(\ell -4)(\ell -6)...(2~~or~~1)$. 

The PSTF part of a tensor, 
\begin{eqnarray}
T_{\la ab \ra}=\left[ {T_{ab}}\right] ^{PSTF}=\left[ {\
\left[ {T_{ab}}\right] ^{STF}}\right] ^P,
\end{eqnarray}
can be constructed recursively from a vector basis following EMT and a
little algebra.

We take the PSTF part of $e^{\la A_\ell \ra}$ \cite
{P,T1,T2,ETMa} to find
\begin{eqnarray}
O^{A_\ell }=e^{\la A_\ell \ra}=\sum_{k=0}^{[\ell /2]}
B_{\ell k}h^{(A_{2k}}e^{A_{\ell -2k})}, \label{II-3}
\end{eqnarray}
where $h^{(A_{2k}}e^{A_{\ell-2k})}\equiv
h^{(a_1a_2}...h^{a_{2k-1}a_{2k}}e^{a_{2k+1}}e^{a_{2k+2}}...e^{a_\ell)}$
and $B_{\ell k}$ are given by (\ref{B}).

{}From \cite{ETMa} we can now construct recursion relations that
play a key role later on. First, 
\begin{eqnarray}
O^{\la A_\ell }e^{a_{\ell +1}\ra }=O^{(A_\ell }e^{a_{\ell +1})}-{\frac
\ell {{2\ell +1}}}e^ah_{ad}O^{d(A_{\ell -1}}h^{a_\ell a_{\ell +1})}
\label{recursion_01} 
\end{eqnarray}
{}From (\ref{recursion_01}), (\ref{defO}), and using 
\begin{eqnarray}
e_{a_1}O^{A_\ell }={\frac \ell {(2\ell -1)}}O^{A_{\ell -1}}, \label{eO-rec}
\end{eqnarray}
it can then be shown that 
\begin{eqnarray}
O^{A_{\ell +1}}=e^{(a_{\ell +1}}O^{A_\ell )}-{\frac{\ell ^2}{(2\ell
+1)(2\ell -1)}} h^{(a_{\ell +1}a_\ell }O^{A_{\ell -1})}  \label{PSTF_rec}
\end{eqnarray}
relates the $(\ell +1)-th$ term to the $\ell -th$ term and the $(\ell
-1)-th$ term. 

The orthogonality, addition theorem and double integral relations
of $O^{A_{\ell}}$ are listed in appendix B. Using the orthogonality
relations we obtain the inversion of the harmonic expansion :
\begin{equation}
\tau (x^i,e^a)=\sum_{\ell =0}^\infty \tau _{A_\ell }(x^i)\,O^{A_\ell
}~\Leftrightarrow ~\tau _{A_\ell }(x^i)=\Delta _\ell ^{-1}\int_{4\pi }d\Omega 
\,O_{A_\ell }\,\tau(x^i,e^a)\,.  \label{inverse} 
\end{equation}

The polynomial $L_{\ell}\equiv O^{A_\ell }O^{\prime }{}_{A_\ell }
=\sum_{m=0}^{[\ell/2]}B_{\ell m}X^{\ell -2m}, $ 
(see (\ref{add01a})) is the natural polynomial
that arises in the PSTF tensor approach (equivalent to the Legendre 
polynomials, see below), where the coefficients $B_{lm}$ are defined by
(\ref{B}). It follows from this that 
\begin{eqnarray}
L_\ell (1)=O^{A_\ell }O_{A_\ell }=\sum_{m=0}^{[\ell /2]}B_{\ell
m}=:\beta _\ell\,~~\mbox{with} 
~~\beta _\ell =\left( {\frac{(\ell !)^22^\ell }{(2\ell )!}}\right)
={\frac{\ell !}{(2\ell -1)!!}}. 
\label{add02a_beta}
\end{eqnarray}
The $\beta _\ell $'s satisfy the recursive relations 
\begin{eqnarray}
\beta _\ell ={\frac{(2\ell +1)}{(\ell +1)}}\beta _{\ell +1},~~\beta
_\ell ={\frac \ell {(2\ell -1)}} \beta _{\ell -1},~~{\frac{\beta
_{\ell +1}}{\beta _{\ell-1}}}={\frac{\ell (\ell +1)}{(2\ell +1)(2\ell
-1)}}.  \label{rec}   
\end{eqnarray}

Any function $W(X)$ can be expanded in terms of the polynomials $L_\ell(X)$ 
\footnote{We include the dipole, $\ell=1$, however the monopole is
dropped as we will only be considering the expansion of the CGI 
perturbations (\ref{3.11}) where the isotropic
part has been factored out according to (\ref{3.10}). We must beware
that this does not cause problems later by omitting the spatial
gradients of the isotropic term.} and then upon combining
(\ref{add01a}) and expansion in terms of $L_{\ell}(X)$ to find 
the expansion in terms $O^{A_\ell}$  : 
\begin{eqnarray}
W(X)=\sum_{\ell=1}^\infty \hat{C}_\ell L_\ell(X)\, 
~~\mbox{and}~~
W(X)=\sum_{n=1}^\infty \hat{C}_nO^{A_n}O^{\prime }{}_{A_n}\,.
\label{add02_add03}
\end{eqnarray}
When $W(X)$ is the angular correlation function, the $\hat{C}_\ell$ are the
corresponding angular power spectrum coefficients (see below). 

\subsubsection{Relationship to Legendre polynomials}

A Legendre polynomial $P_\ell (X)$ is given by renormalising the
polynomials $L_\ell (X)$ defined in (\ref{add01a}) so that $P_\ell
(1)=1.$ By (\ref{add02a_beta}), this implies  
\begin{equation}
P_\ell (X)=(\beta _\ell )^{-1}L_\ell (X)~~\Rightarrow P_\ell (1)=1\,;
\label{pm(x)} 
\end{equation}
consequently from (\ref{add01a}), 
\begin{eqnarray}
O^{A_\ell }O^{\prime }{}_{A_\ell }=\beta _\ell P_\ell (X),  \label{add}
\end{eqnarray}
where $\beta _\ell $ are given by (\ref{add02a_beta}). 
It follows from (\ref{add}) that 
\begin{eqnarray}
P_{\ell}(X) = \sum_{m=0}^{[\ell /2]} A_{\ell m} X^{\ell -2m},
~~\mbox{with}~~ 
A_{\ell k} = {\frac{(-1)^k (2\ell  -2k)! }{2^\ell  k! (\ell -k)! (\ell
-2k)!}}  \label{legdef_II-2} 
\end{eqnarray}
are related to the $B_{lk}$ in (\ref{add1}) by 
\begin{eqnarray}
B_{\ell k} = \left( {\frac{(\ell !)^2 2^\ell  }{(2\ell )!}} \right)
A_{\ell k}  \label{II-6} 
\end{eqnarray}

Any function $W(X)$ can be expanded in terms of both sets of polynomials -
see (\ref{add02_add03}) and the corresponding expression 
\begin{eqnarray}
W(X) &=& \sum_{\ell =1}^{\infty} {\frac{2\ell +1 }{4 \pi}} C_\ell  P_\ell (X).
\label{add01}
\end{eqnarray}
These two expansions can then be related as follows: equating
(\ref{add02_add03}) and (\ref{add01}), and using (\ref{legdef_II-2})
and ({\ref{II-3}), gives  
\begin{eqnarray}
\sum_{\ell =1}^{\infty} \sum_{m=0}^{[\ell /2]} \hat C_\ell  B_{\ell m}
X^{\ell -2m} &=& \sum_{\ell =1}^{\infty} \sum_{m=0}^{[\ell /2]}
{\frac{(2\ell +1) }{4 \pi}} 
 C_{\ell}
 A_{\ell m} X^{\ell -2m}\, ~\Rightarrow
~\hat C_\ell  B_{\ell m} = {\frac{(2\ell +1) }{4 \pi}} C_\ell  A_{\ell m}, 
\end{eqnarray}
from (\ref{II-6}) this gives the relation between the expansion
coefficients :
\begin{eqnarray}
\hat C_\ell  = \left[ {\frac{(2\ell +1) (2\ell )! }{4 \pi 2^\ell
(\ell !)^2}} \right] C_\ell  = \Delta_\ell ^{-1} C_\ell  \,.  
\label{II-25}
\end{eqnarray}

\subsection{The mean square of PSTF coefficients :
 $\la {F_{A_\ell} F^{A_\ell}} \ra$}

It is known as before, from evaluating $\int d\Omega f^2$ and constructing
the orthogonality conditions on $O_{A_\ell}$, that inversion 
\begin{eqnarray}
{F}_{A_\ell}=\Delta _\ell^{-1}\int_{4\pi }d\Omega
O_{A_\ell}f(x^i,e^a)~~~\Leftrightarrow ~~~~f(x^i,e^a)=\sum_{\ell=0}^\infty 
{F}_{A_\ell}O^{A_\ell},
\end{eqnarray}
can be constructed. From this we can build 
\begin{eqnarray}
{F}_{A_\ell} {F}^{B_m}=\Delta _\ell^{-1}\Delta _m^{-1}\int_{4\pi }d\Omega
O_{A_\ell}f(x^i,e^a)\int_{4\pi }d\Omega ^{\prime }{}O^{\prime
}{}^{B_m}f(x^i,e^{\prime }{}^a),
\end{eqnarray}
to find 
\begin{eqnarray}
 {F}_{A_\ell} {F}^{B_m}=\Delta _\ell^{-1}\Delta _m^{-1}\int \int d\Omega
d\Omega ^{\prime }{}O_{A_\ell}O^{\prime }{}^{B_m}f(x^i,e^a)f(x^i,e^{\prime
}{}^a)\,. 
\end{eqnarray}
Taking the ensemble average \cite{LW,W83} then gives 
\begin{eqnarray}
\la { {F}_{A_\ell} {F}^{B_m}} \ra =\Delta
_\ell^{-1}\Delta _m^{-1}\int \int d\Omega d\Omega ^{\prime
}{}O_{A_\ell}O^{\prime}{}^{B_m} \la {f\cdot f^{\prime
}{}} \ra   \label{II-29} 
\end{eqnarray}
where $\la ... \ra $ indicates an ensemble average over
sufficiently many realizations of the angular correlation function.

In order to evaluate this further we assume that the correlations between
the function $f(x^i,e^a)$, {\it {\ i.e. }}, $\la
{f(x^i,e^a)f(x^i,e^{\prime }{}^a)} \ra$ are a function of the angular
separation between the two directions only,  
\begin{eqnarray}
\la {f\cdot f^{\prime }{}} \ra =W(e^ae^{\prime
}{}_a)=W(X)\,. 
\end{eqnarray}
This is a consequence of the Gaussian assumption (\ref{w}), which allows
one to evaluate the angular correlation functions (\ref{ang_correl}) in
a straightforward way \footnote{Ideally one would prefer to evaluate
the angular correlation function without using the Gaussian
assumption, as this is one of the features one should test rather than
assume.}. With this assumption (\ref{II-29}) becomes 
\begin{eqnarray}
\la { {F}_{A_\ell} {F}^{B_m}}\ra =\Delta
_\ell^{-1}\Delta _m^{-1}\int \int d\Omega d\Omega ^{\prime
}{}O_{A_\ell}O^{\prime}{}^{B_m}W(X)\,.  \label{II-29a} 
\end{eqnarray}
Substituting (\ref{add02_add03}) into (\ref{II-29a}), we find 
\begin{eqnarray}
\la { {F}_{A_\ell} {F}^{B_m}} \ra =\Delta
_\ell^{-1}\Delta _m^{-1}\int \int d\Omega d\Omega ^{\prime
}{}O_{A_\ell}O^{\prime}{}^{B_m}\sum_{n=1}^\infty
\hat{C}_nO^{A_n}O^{\prime }{}_{A_n}\,. 
\end{eqnarray}
Rearranging terms, 
\begin{eqnarray}
\la { {F}_{A_\ell} {F}^{B_m}} \ra =\Delta
_\ell^{-1}\Delta _m^{-1}\sum_{n=1}^\infty \hat{C}_n\left\{ {\ \int_{4\pi
}d\Omega O_{A_\ell}O^{A_n}}\right\} \left\{ {\int_{4\pi }d\Omega ^{\prime
}{}O^{\prime }{}^{B_m}O^{\prime }{}_{A_n}}\right\} ,
\end{eqnarray}
where the integrals can be evaluated using the orthogonality conditions on
the $O_{A_\ell}$'s, (\ref{Delta-ortho0-01}), 
\begin{eqnarray}
\la { {F}_{A_\ell} {F}^{B_m}} \ra  &=&\Delta
_\ell^{-1}\Delta _m^{-1}\sum_{n=1}^\infty \hat{C}_n\left\{ {\delta
^{\ell n}\Delta 
_\ell h_{\la A_\ell \ra }{}^{\la A_n \ra }}\right\} \left\{ {\delta ^{mn}\Delta
_m^{+1}h^{\la B_m \ra }{}_{\la A_n \ra }}\right\}   \nonumber \\ 
&=&\sum_{n=1}^\infty \hat{C}_n\delta ^{\ell n}\delta
^{mn}h_{\la A_\ell \ra }{}^{\la A_n \ra }h^{\la B_m \ra}{}_{\la A_n \ra}. 
\end{eqnarray}
Thus 
\begin{equation}
\la { {F}_{A_\ell} {F}^{B_m}} \ra = \hat{C}_\ell\delta
^{\ell m}h_{\la A_\ell\ra }{}^{\la B_m\ra }.
~~\mbox{to find}~~\la { {F}_{A_\ell} {F}^{A_\ell}} \ra 
=\hat{C}_\ell h_{\la A_\ell\ra }{}^{\la A_\ell \ra},
\end{equation}
so using $h_{\la A_\ell\ra }{}^{\la A_\ell\ra}$ (\ref{pstfh-contract})
the mean-square is found to be 
\begin{equation}
\la { {F}_{A_\ell} {F}^{A_\ell}} \ra =\hat{C}_\ell 
\,(2\ell+1) \label{II-36}
\end{equation}
giving the angular power spectrum coefficients $\hat{C}_\ell$ in terms of the
ensemble-averages of the harmonic coefficients \footnote{This corrects
an error in \cite{SAG}, removing a spurious factor of
$3^\ell=h^{A_\ell}_{A_\ell}$ which follows from the orthogonality
conditions in \cite{ETMa} which are corrected here.}. If we use the
Legendre expansion (\ref{add01}) instead of the covariant expansion
coefficients (\ref{add02_add03}), then from (\ref{II-36}) and (\ref{II-25})
the relation is    
\begin{equation}
\la {F_{A_\ell}F^{A_\ell}} \ra =(2\ell+1)\Delta
_\ell^{-1}C_\ell,\label{87} 
\end{equation}
where $C_\ell$ are the usual Legendre angular power spectrum coefficients.

\subsection{The CGI angular correlation function}

We can now gather the results above in terms of the application we have in
mind, namely anisotropy of the CBR. Consider on-sky perturbations made of
Gaussian Random Fields : The {\it angular correlation function
}$C(\beta )$ $=W(X)$ is given by (\ref{w}); the {\it angular power 
spectrum coefficients} $C_\ell$ are given by the mean-square of the 
$\ell-$th temperature coefficient through (\ref{87}): 
\begin{equation}
\la {\tau _{A_\ell}\tau ^{A_\ell}} \ra =(2\ell+1) \Delta_\ell^{-1}C_\ell
\label{129}
\end{equation}
where the constants $\Delta _\ell^{}$ are given by
(\ref{Delta-ortho0-01}). These quantities are related by (\ref{add01}): 
\begin{equation}
W(X)=\sum_{m=1}^\infty C_m{\frac{(2m+1)}{4\pi }}P_m(X)  \label{128}
\end{equation}
where the $P_m(X)$ are given by (\ref{pm(x)}) from (\ref{add01})
and (\ref{add02a_beta}).

\subsubsection{Cosmic variance}

The observations are in fact of $a_\ell ^2$ (which is 
$\sum_{m=-\ell }^{+\ell }|a_{\ell m}|^2/4\pi $ in the usual
notation). This is what is effectively found from experiments, such as
the COBE-DMR experiment. This is a {\it single} realization of the
angular power spectrum $C_\ell $. The finite sampling of events
generated by random processes (in this case Gaussian random fields)
leads to an intrinsic uncertainty in the variance even in 
perfect experiments - this is {\it sample variance}, or in the
cosmological setting, {\it cosmic variance}. We are measuring a single
realization of a process that is assumed to be random; there is an error
associated with how we fit the single realization to the {\it averaged}
angular power spectrum.

The quantities $\la \tau _{A_\ell } \tau ^{A_\ell}\ra$ 
represent the averaged (over the entire ensemble of
possible $C_\ell$'s) angular power spectrum, this is what one is in fact
dealing with in the theory, as the reductions are done in terms of Gaussian
Random Fields where the entire ensemble is considered rather than a single
experimental realization. The $a_\ell^2$ are a sum of the $2\ell+1$ Gaussian
Random Variables $a_{\ell m}$, this is taken to be $\chi ^2$
distributed with $2 \ell+1$ degrees of freedom. Each multipole has 
$2 \ell +1$ samples{\footnote{The uncertainty in $C_\ell$ as 
${\frac{\Delta C_\ell }{C_\ell }} = \sqrt{\frac{2 }{(2 \ell+1)}}$}}.

The key point here is that
 cosmic variance is proportional to $\ell^{-1/2}$ and so is
less significant for smaller angular scales than larger scales (as is
popular wisdom), i.e. {\it cosmic variance} is not an issue on
small scales. 
However on small (and perhaps intermediate scales) systematic errors 
could be underestimated. 

Physical process deviations and instrument noise are expected
to dominate the small scales rather than non-Gaussian effects in the
primordial perturbations, 
but on large
scales the uncertainty due to cosmic variance would swamp out a non-Gaussian
signature. It then seems plausible that on both large and small scales the
assumption of Gaussian Perturbations is acceptable; however on intermediate
scales this is not the case, on these scales the effects of comic variance
would be small enough to allow a non-Gaussian signature to be apparent.
 

\section{Mode expansions}


We now consider spatial harmonic analysis of the angular coefficients
discussed in the previous section. Note that the relations in this section
hold in space-like {\it surfaces}, namely the background space-like surfaces
in an almost-FL model. The application in the following sections will be to
the projection into these spacelike surfaces of null cone coordinates
associated with the propagation of the CBR down the null cone.

Following the Wilson-Silk approach \cite{W83,GSS,HS95a} we consider
the following CGI expansions. Eigenfunctions $Q(x^\nu )$ are chosen to
satisfy the Helmholtz equation 
\begin{equation}
\D^a\D_aQ=-k_{phys}^2 Q \label{eigen}
\end{equation}
in the (background) space sections of the given space-time of interest,
where the $Q$'s are time-independent scalar functions with the physical
wavenumber $k_{phys}(t)=k/a(t)$, the wave number $k$ being independent of
time \footnote{The function $Q$ will be associated with a direction
vector $e^a_{(k)}$ and wave vector $k_a = k e_a^{(k)}$ normal to the
surfaces $Q = \mbox{const}$, see the following subsection.}. These
define tensors $Q_{A_\ell}(k^\nu ,x^i)$ that are Projected,
Symmetric, and Trace-Free, and in the case of scalar perturbations are
chosen to be given by PSTF covariant derivatives of the eigenfunctions $Q$: 
\begin{equation}
Q_{A_\ell }=(-k_{phys} )^{-\ell }\D_{\la A_\ell \ra }Q  \label{def1}
\end{equation}
Using these we define functions of direction and position: 
\begin{eqnarray}
G_\ell [Q](x^\nu ,e^a)=O^{A_\ell}Q_{A_\ell}
\end{eqnarray}
with the $O^{A_\ell }$ defined by (\ref{defO}). Here the $G_\ell $
are call {\it mode operators} and the objects $G_\ell [Q]$ are called
{\it mode functions \footnote{Note these are functions in phase space,
not on $M$.}}.  
It follows that 
\begin{eqnarray}
G_\ell [Q]=(-k_{phys} )^{-\ell }O^{A_\ell }\D_{\la A_\ell \ra }Q  \label{mf}
\end{eqnarray}
and we can expand a given function $f(x^i,e^a)$ in terms of these functions.
In our case this serves as a way of harmonically analysing the
coefficients $\tau _{A_\ell }(x^i)$ in (\ref{3.11}) and (\ref{3.11b}):
expanding the temperature anisotropy in terms of the mode functions, 
\begin{equation}
\tau (x^i,e^a)=\sum_\ell \,\sum_k\,\tau _\ell (t,k)\,G_\ell [Q]
\label{exp} 
\end{equation}
where the $\ell $-summation is the angular harmonic expansion and the
$k$-summation the spatial harmonic expansion (in fact $k$ will be a
3-vector because space is 3-dimensional, see below). Using the
expansion (\ref{3.11}) on the left and (\ref{mf}) on the right, 
\begin{equation}
\sum_\ell \,\tau _{A_\ell}(x^i)\,O^{A_\ell}~=~\sum_\ell
\,\sum_k\,\tau _\ell (t,k)\,(-k_{phys})^{-\ell }O^{A_\ell }\,\D_{\la 
A_\ell \ra }Q 
\end{equation}
and so 
\begin{equation}
\tau _{A_\ell }(x^i)~=~\sum_k\,\tau _\ell (t,k)\,(-k_{phys} )^{-\ell }\D_{\la
A_\ell \ra }Q(x^\nu ) \label{mode_exp}
\end{equation}
which is the spatial harmonic expansion of the radiation anisotropy
coefficients in terms of the symmetric, trace-free spatial derivatives of
the harmonic function $Q$. The quantities $\,\tau _\ell (t,k)$ are the
corresponding mode coefficients. Note that we have not as yet restricted the
geometry of the $Q^{\prime }$s: they could be either spherical or plane-wave
harmonics, for example.

By successively applying the background 3-space Ricci identity, 
\begin{equation}
\D_{abA_\ell }Q-\D_{baA_\ell }Q=+\sum_{n=1}^\ell {\frac K{a^2}}\left( {\delta
_b^{b_n}h_{a_na}-\delta _a^{b_n}h_{a_nb}}\right) \D_{\bar{A_\ell }}Q,
\end{equation}
where $\bar{A_\ell }=a_1...a_{n-1}b_na_{n+1}...a_\ell$, {\it {\ i.e. }}, the
sequence of $\ell $ indices with the $n$-th one replaced with a
contraction. First, the {\it curvature-modified Helmholtz equation} is found
\footnote{This will be necessary in order to switch from the almost-FLRW
multipole divergence equations in part II to the mode representation
where we use (\ref{eO-rec}) to construct $\D^a \D_{\la a}Q_{A_{\ell} \ra}$.}
(using \ref{rel-004},\ref{rel-005},\ref{rel-003}) :
\begin{equation}
\D^a \D_a Q_{\la A_{\ell} \ra} = - \tilde k_{\ell}^2 Q_{\la A_\ell
\ra}~~\mbox{with}~~ - \tilde k_{\ell}^2 = \frac{1}{a^2} \left( {K \ell
(\ell + 2) - k^2} \right)\,. \label{cmhe}
\end{equation}
Second, we are able to construct the {\it mode recursion relation} (using
\ref{rel-001}, \ref{rel-002} in \ref{PSTF_rec}) :
\begin{equation}
e^a\D_a[G_\ell [Q]]=+k_{phys} \left[ {{\frac{\ell ^2}{(2\ell +1)(2\ell
-1)}}\left( {1-{\frac K{k^2}}(\ell ^2-1)}\right) G_{\ell
-1}[Q]-G_{\ell +1}[Q]}\right] , \label{Gl-recursion}
\end{equation}
The latter is the basis of the standard derivation of the 
linear-FRW mode heirarchy for scalar modes. The derivation
of these are given in appendix C (we consider only scalar 
eigenfunctions). It will be seen, in paper II, that this relation can 
be used in place of the general divergence relations which allow the 
construction of generic multipole divergence equations \cite{ETMa} if 
one restricts oneself to constant curvature space-times.

Given the recursion relation one can immediatly make the connection
with the usual Legendre tensor treatment (\ref{len-PSTF}) this is
shown in appendix C (\ref{len-PSTF}-\ref{len-rec}).


\subsection{$|\tau _\ell|^2$ in almost-FLRW universes}

We now relate the multipole mean-squares $\left< \tau_{A_\ell} \tau^{A_\ell}
\right>$ of the ensemble average over the multipole moments with that of the
mode coefficient mean-squares $|\tau_\ell(k)|^2$.

In order to carry this out we relate two separate spatial harmonic
expansions (\ref{mode_exp}) for the same function: the first is one
associated with plane wave harmonics ($Q^k$), naturally used in describing
structure existing at any time $t$, and the second, one associated with
radial and multipole harmonics ($O^{A_\ell}_{(\chi)}$ $R_{A_\ell}$), 
{\it {\ i.e. }}, a spherical expansion based at the point of
observation, naturally arises when we project the null cone angular
harmonics into a surface of constant time. These are both related to
the Mode Function formulation which becomes useful in the non-flat
constant curvature cases. 


\subsubsection{Plane-waves and Mode functions}


Considering flat FRW universes, each set of eigenfunctions
satisfy (\ref{eigen}). The temperature anisotropy (\ref{3.10}) 
can be expressed in terms of its plane-wave spatial Fourier
transform :
\begin{equation}
\tau(x^i,e^a) = \sum_{k_{\nu}} \tau(k,t,e_a^{(k)},e^a) Q_{flat}.\label{t-pw}
\end{equation}
(For a more detailed treatment see appendix D). It can be
shown that for the flat case, $K=0$, that is (\ref{q-defs}) hold
in (\ref{q_a}) and (\ref{q-eigen}) to find from (\ref{plane-wave}) :
\begin{equation}
\D_{\la A_{\ell} \ra} Q_{flat} = (- i k_{phys})^{\ell}
O_{A_{\ell}}^{(k)} Q_{flat}  \label{DQ-pw}
\end{equation}
where $O_{A_{\ell}}^{(k)}$ are the PSTF tensors associated with the
direction $e^a_{(k)}$. Then from (\ref{def1}) we find that :
\begin{equation}
Q_{flat}{}_{A_\ell} = (-1)^{\ell} O_{A_\ell}^{(k)} Q_{flat}, \label{QAl-pw}
\end{equation}   
along with (\ref{mode_exp}) to get the temperature multipole :
\begin{equation}
\tau_{A_{\ell}} = (-1)^{\ell} \sum_k \tau_{\ell}(t,k) Q_{flat} 
O_{A_\ell}^{(k)}. \label{tAl-pw}
\end{equation}


\subsubsection{Radial expansions and Mode functions}


Using the flat, $K=0$, spherical eigenfunctions centred on a
point $x^i_0$, and with associated radial direction vector
$e_a^{(\chi)}$. The latter is the same as the (spherically symmetric)
projection into the constant time surfaces of the tangent vector $e^a$
of the radial null geodesics, so we need not distinguish it from that
vector. In this case the $\ell$-th harmonic is  
\begin{equation}
Q_\ell (x^\alpha,e^a_{(\chi)})|_{\chi} = R_{A_\ell }(r) O^{A_\ell
}_\chi, ~~\D_a r = e_a, ~~\D^a e_a = {\frac{2 }{r}}  \label{sph1}
\end{equation}
where $e^a = dx^a/dr$ is the unit radial vector. Cartesian coordinates in
space are given by $r$ and $e^a$ through $x^i = r e^i$. Defining the
projection tensor 
\begin{equation}
p_{ab} = h_{ab} - e_a e_b \Rightarrow p_{ab} u^a =0= p_{ab}
e^a,~~p_a^{~a}=2,~~p^a_{~b} p^b_{~c} = p^a_{~c},
\end{equation}
then ($e^a$ is shear and curl free) 
\begin{eqnarray}
\D_a e_b = \frac{1}{r} p_{ab} \Rightarrow \D_a p_{bc} = - \frac{1}{r} (p_{ab}
e_c + p_{ac} e_b), ~~ \D^a p^b_{~a} = - \frac{2}{r}e^b
\end{eqnarray}
so from (\ref{eigen}) 
\begin{equation}
\D_a \D^a (R_{A_\ell } O^{A_\ell }) = - k_{phys}^2 (R_{A_\ell } O^{A_\ell }).
\end{equation}
To work out the {\it l.h.s.}, we must first calculate 
$\D_a Q_{\ell}$ (\ref{DQ-l}) and $\D_a \D^a Q_{\ell}$
(\ref{D2Q-l}). These are then used to calculated :
$\D_a R_{A_\ell}$ (\ref{DRAl}), $\D^a \D_a R_{A_\ell}$ (\ref{D2RAl}),
$\D_a R^{A_\ell}$ (\ref{DOAl}) and $\D_a \D^a O^{A_\ell}$
(\ref{D2OAl}) respectively.

Putting these in (\ref{D2Q-l}) to find (\ref{proto-bessel}) frow which 
we get :
\begin{eqnarray}
\frac{1}{r^2} {\frac{\p  }{\p  r}} \left( r^2 {\frac{\p R_{A_\ell }
}{\p  r}} \right) + R_{A_\ell } \left({- \frac{\ell (\ell +1)}{r^2}}
\right) = - k_{phys}^2 R_{A_\ell }\,,  \label{bessel} 
\end{eqnarray}
the spherical Bessel equation. Thus providing the PSTF derivation of
the spherical bessel equation in terms of the irreducible representation.

Now consider that we can choose any basis we like for the tensor basis here,
independent of the spatial coordinates used. It is convenient to use the
plane wave decomposition to get a parallel vector basis. We do this by
writing 
\begin{eqnarray}
R_{A_\ell }(r) = \sum_{k^a} R_{A_\ell }^{(k)} = \sum_{k^a} R_\ell
(k,r) O_{A_\ell }^{(k)}\,. \label{103}
\end{eqnarray}
this expresses the tensor eigenfunction in terms of the monopole
eigenfunction. When this is substituted into (\ref{proto-bessel}) we
obtain the {\it radial equation} which has solutions that are
{\it spherical Bessel functions} in the flat case :
\begin{eqnarray}
{\frac{\p  }{\p  r}} \left( {r^2 {\frac{\p  R_{\ell } }{\p  r}}}
\right) + \left[ {k_{phys}^2 r^2- \ell (\ell +1)} \right] R_\ell  = 0\,,
~~\mbox{to find}~~ R_\ell (r) = 
\alpha_l\, j_\ell (k_{phys} r)  \label{sph-bessel}
\end{eqnarray}
where $k^2_{phys} = k^2/a^2$ and $\alpha_\ell $ are integration constants (the
second set of constants for this second order equation vanish because we
choose $R_l(0)$ to be finite; the Neumann functions are not finite at $r=0$).

{}From (\ref{sph1}), (\ref{103}), and (\ref{sph-bessel}) we have found
that the solutions to the Helmholtz equations give the eigenfunctions 
\begin{eqnarray}
Q_\ell (x^\nu) &=& \alpha_\ell \, \sum_{k^{\nu}} j_\ell (k_{phys} r)
O^{A_\ell }_{(k)} O_{A_\ell }^{(\chi)} = \alpha_\ell \, \sum_{k^{\nu}}
j_\ell (k_{phys}(k,r) r) L_\ell (X). 
\end{eqnarray}
and we can set $\alpha_\ell  = (\Delta_\ell )^{-1}$ so that $\int
Q_\ell  d\Omega = 1.$ 

It is important to notice that the functions $L_\ell (X)$ depend both 
on $k^a$ and on $e^a$, and so for each $k^a$ is a function of 
$(\theta,\phi)$, thus $Q$ is indeed a function of all
spatial coordinates. We can pick any direction $k^a$ to find the particular
eigenfunctions 
\begin{eqnarray}
Q_\ell ^{(k)} = (\Delta_\ell )^{-1} j_\ell (k_{phys} r) O^{A_\ell
}_{(k)} O_{A_\ell }^{(\chi)}. 
\end{eqnarray}
associated with that direction. The general $\ell $-th eigenfunction
is a sum of such eigenfunctions over a basis of directions $k^a$
\footnote{Many treatments choose a particular direction for $k$: $k^a
= \delta^a_3$ or similar and omit the summation.}:  
\begin{equation}
Q_\ell (x^\nu) = \sum_{k^a} Q_\ell ^{(k)} \,.
\end{equation}
\noindent
Now we find $O^{A_{\ell}} D_{A_{\ell}}Q$ (\ref{QD-Rl}) in terms of $R_{\ell}$
(as show in the appendix using PSTF techniques, \ref{104a}-\ref{QD-Rl}) :
\begin{equation}
O^{A_\ell }_{(\chi)} \D_{\la A_\ell \ra } Q = (- k_{phys} )^{\ell }
O^{A_\ell }_{(\chi)} O_{A_\ell }^{(k)} R_\ell  ~~\Rightarrow ~~
O^{A_\ell } Q_{A_\ell } = (- k_{phys})^\ell  Q_\ell. 
\end{equation}
Putting this in the expansion (\ref{mode_exp}) 
\begin{eqnarray}
\tau_{A_\ell }(x^i) = \sum_{k^{\nu}} \tau_\ell (t,k)
(-k_{phys})^{-\ell } \D_{\la A_\ell \ra } Q(x^{\nu}) = \sum_{k^{\nu}}
\sum_n \tau_\ell (t,k) (-k_{phys})^{-\ell } \D_{\la A_\ell \ra } Q_n
\label{mean-s2} 
\end{eqnarray}
gives the present version of (\ref{exp}) 
\begin{equation}
\tau(x^i,e^a_{(\chi)}) = \sum_\ell  \tau_{A_\ell } O^{A_\ell } =
\sum_\ell  \tau_\ell (k,t) G_\ell [Q];
\end{equation}
directly analysing the coefficients $\tau_{A_\ell }$ in terms of these
functions $Q$ - given that we have 3-dimensions worth of variability
so as to represent arbitrary spatial functions - with purely
time-dependent co-efficients (parametrized by a vector $k^a$),  


\subsubsection{Radial expansions and plane-waves}


Now consider the inversion\footnote{Contrast with (\ref{sph1}):
there the l.h.s is the spherical eigenfunction; here it is the plane
one, expressed in terms of spherical ones.} 
\begin{equation}
e^{(+i k_{\alpha} x^{\alpha})} = \sum_{\ell } \tilde{R}_{A_\ell }(r)
O^{A_\ell }_{(\chi)},~~\iff~~ \tilde{R}_{A_\ell } = \Delta_\ell ^{-1} \int d
\Omega_{(\chi)} e^{+i k_{\alpha} x^{\alpha}} O_{A_\ell }^{(k)}
\end{equation}
by taking a Taylor expansion using $k_{\alpha} x^{\alpha} = k_{phys} r
e_b^{(k)} e^{b}_{(\chi)}$ we find 
\begin{eqnarray}
\tilde{R}_{A_\ell } &=& \Delta_\ell ^{-1} \sum_{n=0}^{\infty}
{\frac{(+i)^n (k_{phys} r)^n }{n!}} e_{B_n}^{(k)} \int d \Omega
O_{A_\ell }^{(\chi)} e^{B_n}_{(\chi)}  \nonumber \\
&=& \Delta_\ell ^{-1} \sum_{n=0}^{\infty} {\frac{(+i)^n (k_{phys} r)^n
}{n!}} {\frac{\delta_n^{\ell +2m} n! (n-\ell +1)!! }{(n-\ell +1)!
(n+\ell +1)!!}} 4 \pi e_{B_n}^{(k)} h_{\la A_\ell \ra }{}^{\la B_\ell
} h^{B_{n-\ell }\ra }.  \label{p-r003} 
\end{eqnarray}
on using (\ref{dint}). Putting (\ref{p-r004}) into (\ref{p-r003})
we find 
\begin{eqnarray}
\tilde{R}_{A_\ell } &=& (+i)^\ell  \Delta_\ell ^{-1} 4 \pi
\sum_{m=0}^{\infty} {\frac{(-1)^m (k_{phys} r)^{\ell +2m} (2m+1)!!
}{(2m+1)! (2m+2\ell +1)!!}} O_{A_\ell }^{(k)}  
\nonumber \\
&=& 4 \pi (+i)^\ell  \Delta_\ell ^{-1} \sum_{m=0}^{\infty}
{\frac{(-1)^m (k_{phys} r)^{\ell +2m} }{2^m m! (2m+2\ell +1)!!}}
O_{A_\ell }^{(k)}. 
\end{eqnarray}
This can be re-expressed as 
\begin{equation}
\tilde{R}_{A_\ell } = 4 \pi (+i)^\ell  \Delta_\ell ^{-1} j_\ell
(k_{phys} r) O_{A_\ell }^{(k)}
= (+i)^\ell  (2\ell +1) \beta_\ell ^{-1} j_\ell
(k_{phys} r) O_{A_\ell }^{(k)}. 
\end{equation}
Hence 
\begin{eqnarray}
Q_{flat} (x^i,e^{(k)}_a)= e^{+i k^{\alpha} x_{\alpha}} = 4 \pi
\sum_{n,k^{\nu}}^{\infty} (+i)^n j_n(k_{phys} r) O_{(k)}^{A_n}
O_{A_n}^{(\chi)} \Delta_n^{-1}  \label{p-rfinal}
\end{eqnarray}
links the plane-waves to the spherical expansion 
This recovers, from $L_\ell (e^a e^{\prime}{}_a) = \beta_\ell  P_\ell (e^a
e^{\prime}{}_a)$ the more usual  
\begin{equation}
e^{+i k_{phys} r X} = \sum_\ell  \tilde{R}_{A_\ell } O^{A_\ell
}{}_{(\chi)} = \sum_\ell (+i)^\ell  (2\ell +1) j_\ell (k_{phys} r)
P_\ell (X) 
\end{equation}


\subsubsection{$|\tau _\ell |^2$ for $K=0$ (flat) almost-FLRW models}


We now return to the relationship between the
$\tau_{A_\ell }$ and  $\tau_\ell $. Now from (\ref{plane-wave}), (\ref{102}),
(\ref{p-rfinal}) and (\ref{mean-s2}) : 
\begin{eqnarray}
\tau_{A_\ell } &=& (+i)^\ell  \int {k^2 d k \over (2 \pi)^3}
 \int d \Omega_k \tau_{\ell }(k,t)
O^{(k)}_{A_\ell } \sum_{n=0}^{\infty} (+i)^n j_n(\lambda r) O_{(k)}^{B_n}
O_{B_n}^{(\chi)} \beta_n^{-1} (2n+1).
\end{eqnarray}
This can be reduced further (\ref{tau_Al_planewv}-\ref{Atau_Al_radial}), 
where upon using (\ref{Rl_defn}) one finds that 
\begin{eqnarray}
\tau_{A_\ell } = 4 \pi O_{A_\ell }^{(\chi)} \int {k^{\prime}{}^2
dk^{\prime}\over (2 \pi)^3 }{}\tau_\ell (k^{\prime}{},t) 
R_\ell (k^{\prime}{},\chi),\label{tau_Al_radial}
\end{eqnarray}
where $\chi=r/a$ and by using $R_\ell (k,\chi)= j_\ell (\lambda r)$ we
can identify this with (\ref{tau_Al_planewv}), the equivalent of the
multipole moments found using the explicit form of the plane-waves.

In order to proceed further, from (\ref{tau_Al_radial}) we construct the
ensemble average
{\footnote{Remembering that 
\[
\la { j_{\ell}( k' r) j_{\ell}(k,r) } \ra_{x^a} = 4 \pi 
\left( {\pi \over 2} \right) {1 \over k^2} \delta(k'-k).
\]
}} 
\begin{eqnarray}
\la  {\tau_{A_\ell } \tau^{A_\ell }} \ra  &=& (4 \pi)^2 O_{A_\ell }^{(\chi)}
O^{A_\ell }_{(\chi)} \left<  {\int {k^{\prime}{}^2 dk^{\prime}\over(2 \pi)^3}
{}\tau_\ell (k^{\prime}{},t) R_\ell (k^{\prime}{},r) \int {k^2 dk \over (2
\pi)^3} \tau_\ell (k,t) R_\ell (k,r) } \right> ,  \nonumber \\ 
&=& {4 \pi \over (2 \pi)^3} \beta_\ell  \int k^2
dk | \tau_\ell (k,t)|^2. 
\end{eqnarray}
What has happened here is that we imagine an ensemble of universes and we
use an ensemble average rather than the space average; we have to do this
because $\tau_\ell $ is not square-integrable, {\it i.e.}, we cannot use the
r.m.s value as we cannot integrate the square over the space in general;
this operation is not well defined. In order to deal with the ensemble
average over the mode coefficients, {\it {\ i.e. }} $\la  \tau_\ell ^k
\tau_m^{k^{\prime}{}} \ra $, we assume the perturbations to be fairly
homogeneously spread throughout the space and not confined in a particular
region, and assume that there are no correlations between perturbations with
different wavenumbers. Here the Gaussian assumption is useful, we have that 
\[
\left<  {\tau_\ell (k,t) \tau_\ell (k^{\prime}{},t)} \right> =
  (2 \pi)^3 \delta(k^{\prime}{}-k) |\tau_{\ell}(k,t)|^2. 
\]


\subsubsection{$|\tau _\ell |^2$ for almost-FLRW models}

It is useful to notice that an alternative although
equivalent avenue of approach is also possible proceeding directly from the
mode expansion in $G_\ell [Q]= O^{A_\ell }_{(\chi)} Q_{A_\ell }$; this can be used for
any $K$.

Notice that as before at the observer at $x_0^i$, 
\begin{equation}
\tau(x_0^i,e^a) = \sum_\ell  \tau_{A_\ell } O^{A_\ell },~~ \iff~~
\tau_{A_\ell }(x_0^i) = \Delta_\ell ^{-1} \int d \Omega O_{A_\ell }
\tau(x_0^i,e^a). 
\end{equation}
Now in the spatial section in general we can write 
\begin{equation}
\tau(x^i,e_{(k)}^a) = \sum_\ell  \tau_{A_\ell } O^{A_\ell }_{(k)}, ~~ \iff ~~
\tau_{A_\ell }(x^i) = \Delta_\ell ^{-1} \int d \Omega^{(k)} O_{A_\ell }^{(k)}
\tau(x^i,e^a_{(k)}).
\end{equation}
Now the point $x_0^{\nu}$ is chosen at some earlier time in a spatial
section, with radial direction vector $e^a_{(\chi)}$, for FRW models we can
consider $e^a_{(\chi)}$ and $e^a$ to be equivalent and write 
\begin{eqnarray}
\tau(x_0^i,e^a) = \tau(x_0^i,e^a_{(\chi)}) = \sum_\ell  \tau_{A_\ell }(x_0^i)
O^{A_\ell }_{(\chi)} \equiv \sum_\ell  \tau_{A_\ell }(x_0^i) O^{A_\ell
}  \nonumber \\ ~~\iff ~~ \tau_{A_\ell }(x_0^i) = \Delta_\ell ^{-1}
\int d \Omega^{(\chi)} O_{A_\ell }^{(\chi)} \tau(x_0^i,e^a_{(\chi)}).
\end{eqnarray}

\noindent
Remember that in some spatial section (where the integral relations
(\ref{OGl},\ref{GlGm},\ref{Gl-norm1}) are useful)
\begin{equation}
\tau(x^i,e^a) = \sum_{\ell ,k^{\nu}} \tau_\ell (k,t) O^{A_\ell }_{(\chi)}
Q_{A_\ell }. 
\end{equation}

Now from 
\begin{equation}
\tau_{A_\ell }(x_0^i) = \Delta_\ell ^{-1} \int d \Omega
\sum_{m,k^{\alpha}} \tau_m O_{A_\ell } G_\ell [Q],
\end{equation}
we identify $O^{A_\ell}_{(\chi)}$ with $O^{A_\ell }$, to find 
\begin{equation}
\tau_{A_\ell } = \sum_{k^{\alpha}} \tau_\ell (k,t) Q_{A_\ell },
\end{equation}
which means that for $K=0$ ($R_\ell  \propto j_\ell $). 
\begin{eqnarray}
\left< {\tau_{A_\ell } \tau^{A_\ell }} \right> &=& \sum_{k^{\nu}}
|\tau_\ell |^2 \la {Q_{A_\ell } Q^{A_\ell} }\ra = (4 \pi) \beta_\ell
\sum_{k} |\tau_\ell |^2. \label{mean-s3}
\end{eqnarray}

What is particularly useful about the last method of calculation is that for
constant $K$ surfaces 

it can be shown on using relation (\ref{rec_norm_Gn}) recursively that 
\begin{equation}
\int d \Omega_k \left< |G_\ell [Q] G_m[Q]| \right> = 4 \pi \Xi_{\ell
}^2 \delta^m_\ell\,, ~~\mbox{with}~~ 
\Xi_\ell ^2 = \prod_{n=1}^\ell  (\alpha_n)^2
\end{equation}
and $\left< |\ldots | \right>$ denotes the ensemble average of the
eigenfunctions. What is useful here is to notice that (using the
normalisation for $K=0$), 
\begin{equation}
\Xi_\ell ^2|_{K=0} = \prod_{n=1}^{\ell } {\frac{n^2 }{(2n+1)(2n-1)}} =
{\frac{\beta_\ell ^2 }{2\ell +1}} = (4 \pi)^{-2} (2\ell +1)
\Delta_\ell ^2. 
\end{equation}
This means that for $K=0$ 
\begin{equation}
\int d \Omega_{(k)} \left< | G_\ell ^k G_m^{k^{\prime}{}} | \right> =
\delta_\ell ^m  {\frac{{2\ell +1} }{4 \pi}} \Delta_\ell ^2.  \label{Gl-norm2}
\end{equation}

Now we can extend the above to FL models with $K\neq 0$. The mean-square for
constant $K$ is found by modifying the normalization after identifying the $%
K=0$ normalizations in (\ref{Gl-norm1}) and (\ref{Gl-norm2}) : 
\begin{equation}
\la  {\tau _{A_\ell }\tau ^{A_\ell }}\ra  = \sum_{k^\nu
}|\tau _\ell |^2 \la {Q_{A_\ell }Q^{A_\ell}} \ra=(4\pi ) \beta _\ell
\sum_k|\tau _\ell |^2\Xi _\ell^2, 
\end{equation}
or alternatively keeping the form of the mean-square in (\ref{mean-s3})
by redefining the mode function expansion \cite{WSc} by a
wavelength-dependent coefficient; then 
\begin{equation}
\tau (x^i,e^a)=\sum_{\ell ,k^\nu }\tilde{\tau}_\ell (k,t)M_\ell [Q]
\,, ~~\mbox{and}~~ 
M_\ell [Q]=\Xi _\ell (k,K)G_\ell [Q], \label{redef}
\end{equation}
defines the new coefficients $\tilde{\tau }_\ell (k,t)$ so that 
\begin{equation}
|\tilde{\tau}_\ell |^2=|\tau _\ell |^2\Xi _\ell ^2\,. \label{prb_208b} 
\end{equation}
Because we have redefined the mode functions in (\ref{redef}), the form
of the equations for $K\neq 0$ is the same as in the case $K=0$. However the
coefficients are different because they are from a different expansion.
Using the results from the $K=0$ case, either (\ref{tau_Al_radial}) or
(\ref{tau_Al_planewv}), we have that 
\begin{eqnarray}
\la  {\tau _{A_\ell }\tau ^{A_\ell }}\ra  = {1 \over 2 \pi^2 }
\beta _\ell \int {dk \over k}k^3 |\tau _\ell (k,t)|^2.  \label{prb_208}
\end{eqnarray}
Hence, on using $\hat{C}_\ell =\Delta ^{-1}C_\ell $ and $\la  { \tau
_{A_\ell }\tau ^{A_\ell }}\ra  =\hat{C}_\ell (2\ell +1)$, we now have 
\begin{eqnarray}
\la  {\tau (x_0^i,e^a)\tau (x_0^i,e^{\prime }{}^a)}\ra 
=\sum_{\ell =0}^\infty (2\ell +1)^{-1}\la  {\tau _{A_\ell }\tau
^{A_\ell }} \ra  O_{B_\ell }O^{\prime }{}^{B_\ell }
\end{eqnarray}
which can now be written from (\ref{prb_208}) and (\ref{prb_208b}) as 
\begin{eqnarray}
\la  {\tau (x_0^i,e^a)\tau (x_0^i,e^{\prime }{}^a)}\ra  
&=& \left( {1 \over 2 \pi^2} \right) \sum_{\ell =0}^\infty 
(2\ell +1)^{-1}\beta _\ell \int_0^\infty {dk \over k} k^3 
|\tilde{\tau}_\ell (k,t)|^2 O_{A_\ell }O^{\prime}{}^{A_\ell}
\nonumber \\ &=& \left( {1 \over 2 \pi^2}\right) 
\sum_{\ell=0}^\infty \beta _\ell^2(2\ell +1)^{-1} \int {dk \over k} k^3 
|\tilde{\tau}_\ell (k,t)|^2 P_\ell (e^ae^{\prime }{}_a)  
\label{prb_208a}
\end{eqnarray}
thus reobtaining the results of White and Wilson :
\begin{eqnarray}
C_{\ell} = \frac{2}{\pi} {\beta_{\ell}^2 \over (2 \ell +1)^2}
\int_{0}^{\infty} {dk \over k} k^3 | \tilde \tau_{\ell}(k,t)|^2.
\end{eqnarray}
One needs to be careful here with the factors of $(2\ell +1)$.
Equation (\ref{prb_208a}) relates the amount of power in a given 
wavenumber, $|\tau_\ell(k,\eta_0)|^2$, given the intersection of the 
null-cone fixed at the observer $x_0^i$, on a angular scale $\ell $ given 
that the angular correlations are found on the scale $X=e^a e^{\prime}{}_a$, 
{\it {\ i.e. }}, X is the separation between measurements.


\section{Conclusions}

We have given here a comprehensive survey of the CGI 
representation of CBR anisotropies in almost-FLRW universes, and
related this formalism to the other major formalisms in use for this purpose
at the present time. 

This paper has been concerned with algebraic relations: specifically
the Multipole, (\eg for $\tau_{A_\ell}$), and Mode, (\eg for $\tau_{\ell}$), 
formalisms and the relationship between these. Where possible the multipole
moments have been treated for a spacetime with generic inhomogeneity and
anisotropy but small temperature anisotropies. The mode moments however
are only meaningful in the restricted class of almost-FLRW universes.

The subsequent papers in the series consider the differential relations
satisfied by the quantities mentioned here \cite{MGE}, and will show how 
both timelike and null integrations are used to lead to the standard results
in the literature. Taken together, this will be is an 
{\it ab initio} demonstration of the way the different formalisms in use, 
and their major results, can  be obtained from a single CGI 
approach, as well as providing the natural extension of the usual 
results into the non-linear (exact) theory.


\section*{Acknowledgements}

We are grateful to Bill Stoeger, Bruce Bassett, Roy Maartens, and 
Peter Dunsby for
useful comments and suggestions. This work was supported by the South 
African Foundation for Research and Development.

\newpage

\appendix

\section{Spherical Harmonics}

\subsection{Basic relations}

A Spherical Harmonic (SH) $Y_{\ell ,m} (\theta, \phi)$ is related to an
Associated Legendre Polynomial (ALP) \cite{T1,P}, 
\begin{eqnarray}
Y^{\ell m} &=& C^{\ell m} e^{im \phi} P^{\ell m}(\cos \theta), \\
&=& C^{\ell m} (e^{i \phi} \sin \theta)^m \sum_{j=0}^{[(\ell -m)/2]}
A^{\ell mj} (\cos \theta)^{\ell -m-2j}~~~~\forall~m \geq 0.
\end{eqnarray}
Here, 
\begin{eqnarray}
C^{\ell m} = (-1)^m {\frac{{(2\ell +1)(\ell -m)!}}{{4 \pi (\ell +m) !}}}\,,
~~\mbox{and}~~
A^{\ell mj} = {\frac{(-1)^j (2\ell  -2j)! }{2^\ell  j! (\ell -j)! 
(\ell -m-2j)!}}, 
\end{eqnarray}
aloong with 
\begin{eqnarray}
Y^{\ell m} = (-1)^m Y^{\ell |m|*}~~~\forall~~ m \leq 0.
\end{eqnarray}
Now we can relate the SH, $Y^{\ell m}$, to the direction vector
product $e^{A_\ell }$,  
\begin{eqnarray}
Y^{\ell m} = {\cal Y}^{\ell m}_{A_\ell } e^{A_\ell }(\theta,\phi),
\end{eqnarray}
where following \cite{P} (from making the substitution $e^1 + i e^2 =
e^{i \phi} \sin \theta$ and $e^3 = \cos \theta$ into the above relation) 
\begin{eqnarray}
{\cal Y}^{\ell m}_{A_\ell } &=& C^{\ell m} \sum_{j=0}^{[(\ell -m)/2]}
A^{\ell mj} \prod_{k=0}^{m} \left( {h^1_{\; (a_k} + i h^2_{\; (a_k}}
\right) \prod_{p={m+1}}^{\ell -2j} h^3_{\; a_p}  \nonumber \\
&\;& ~~~~~\times \prod_{q=1}^{j} (h^{\alpha}_{\; a_{2q -1 + \ell  - 2j}}
h_{\alpha a_{2q + \ell  - 2j})}).
\end{eqnarray}
Furthermore, it can then be shown that from 
\begin{eqnarray}
f = \sum_{\ell } F_{A_\ell } e^{A_\ell }\,,~~\mbox{and}~~
F_{A_\ell } = \sum_{m=-\ell }^{m=+\ell } a^{\ell m} {\cal Y}^{\ell
m}_{A_\ell }. 
\end{eqnarray}
This is not unexpected.

\subsection{Consequences}

\subsubsection{Closure}

\begin{eqnarray}
\sum_{\ell =0}^{\infty} \sum_{\ell =-m}^{\ell =+m} Y^*_{\ell
m}(\Omega^{\prime}{}) Y_{\ell m}(\Omega) = \delta(\Omega-
\Omega^{\prime}{}) . 
\end{eqnarray}

\subsubsection{Addition}

\begin{eqnarray}
\sum_{\ell =-m}^{\ell =+m} Y_{\ell m}(\Omega^{\prime}{}) Y_{\ell
m}(\Omega) = {\frac{(2\ell +1)}{4 \pi}} P_\ell  = \Delta^{-1}_\ell
O_{A_\ell } O'^{A_\ell }. 
\end{eqnarray}

\subsubsection{Orthonormality}

\begin{eqnarray}
\int_{4 \pi} d \Omega Y_{\ell
^{\prime}{}m^{\prime}{}}(\Omega^{\prime}{}) Y_{\ell m} (\Omega) =
\delta_{\ell  \ell ^{\prime}{}} \delta_{m m^{\prime}{}}. 
\end{eqnarray}

\subsubsection{Matching plane-waves to spherical harmonics}

\begin{eqnarray}
e^{(ik\chi )} &=&4\pi \sum_\ell i^\ell j_\ell (k\chi )Y_{\ell
m}Y^{\prime }{}_{\ell ,-m}=4\pi \sum_\ell i^\ell j_\ell \Delta _\ell
^{-1}O^{A_\ell }O^{\prime}{}_{A_\ell }, \\ e^{(ik\chi )} &=&\sum_\ell
i^\ell (2\ell +1)j_\ell P_\ell =\sum_\ell i^\ell (2\ell +1)j_\ell
\beta_\ell ^{-1}O^{A_\ell }O^{\prime }{}_{A_\ell }. 
\end{eqnarray}

\section{Multipole relations}

\subsection{Properties of $e^{A_{\ell}}$}


\subsubsection{Normalization} 

The normalization for $e^{A_{\ell}}$ is found from \cite{ETMa}, for
{\it odd} and {\it even} $\ell$ respectively : 
\begin{eqnarray}
\frac{1}{4 \pi} \int_{4 \pi} e^{A_{2 \ell+1}} d \Omega = 0,
~~\mbox{and}~~  
\frac{1}{4 \pi} \int_{4 \pi} e^{A_{2 \ell}} d \Omega = \frac{1}{2 \ell+1}
h^{(A_{2 \ell})}\,.
\end{eqnarray}
{}From which (contracting with $h_{(A_{2 \ell})}$) it can be shown that 
\begin{equation}
h_{(A_{2 \ell})} h^{~(A_{2 \ell})} = (2 \ell+1)\, \label{contr}
\end{equation}
this can also be shown algebraically \cite{dGvLvW}.
 
\subsubsection{Orthogonality} 

{}From \cite{ETMa} we also have that 
\begin{eqnarray}
\int_{4\pi }e^{A_\ell}e^{B_m}d\Omega ={\frac{4\pi
}{\ell+m+1}}h^{(A_\ell B_m)}.
\label{orth1}
\end{eqnarray}
if $\ell+m$ is even, and is zero otherwise (this follows from the 
above because $e^{A_\ell}e^{B_m}=e^{A_{\ell+m}}$ on relabeling 
indices: $b_1..b_m\rightarrow a_{\ell+1}..a_{\ell+m}$.) 

\subsubsection{Addition Theorem} 

{}From (\ref{3.1}) 
it follows that 
\begin{eqnarray}
e^{A_\ell } e^{\prime}{}_{A_\ell } = (X)^{\ell }\,~~ \Rightarrow~~
\sum_{\ell =0}^{\infty} e^{A_\ell } e^{\prime}{}_{A_\ell } 
= \sum_{\ell } (X)^{\ell } = {\frac{\cos \beta }{{1- \cos \beta}}},
\end{eqnarray}
where $X= \cos \beta$. It may be useful to compare these to the relations
for standard spherical harmonics, which are given in the appendix
A. Note that 
\begin{eqnarray}
\int_{4 \pi} d \Omega e^{A_n} e^{\prime}{}_{A_n} = \int_{4 \pi} d \Omega
(e^a e^{\prime}{}_a)^n = 2 \pi \int_{-1}^{+1} d X X^n = \left\{ 
\begin{array}{cl}
0~~~~\forall ~~n~\mbox{odd} &  \\ 
{\frac{4 \pi }{n+1}} \forall~~n~\mbox{even} & 
\end{array}
\right. ,
\end{eqnarray}
where the integral is taken over $e^a$ with $e^{\prime}{}^a$ fixed. 

\subsubsection{Orthogonality of $O^{A_{\ell}}$}

The orthogonality conditions can be found from 
\begin{eqnarray}
\sum_{m,\ell }\int d\Omega ( {F}_{A_\ell }O^{A_\ell })( {F}_{B_m}O^{B_m}),
\end{eqnarray}
see \cite{ETMa}. Here $F_{A_\ell }$ are arbitary PSTF 
harmonic components of some $f(e^a,x^i)$. Using (\ref{orth1}), 
(\ref{II-3}), and (\ref{B}) we find 
\begin{eqnarray}
\int d\Omega O^{A_\ell }O_{B_m}=\delta _m^l\Delta _\ell h^{\la A_\ell
\ra }{}_{\la B_\ell \ra }~~\mbox{with}~~
\Delta _\ell :={\frac{4\pi }{(2\ell +1)}}{\frac{2^\ell (\ell !)^2}
{(2\ell )!}}.  \label{Delta-ortho0-01}
\end{eqnarray}
where $h^{\la A_\ell \ra }{}_{\la B_\ell \ra }=h^{\la a_1}{}_{\la
b_1}\cdots h^{a_\ell \ra }{}_{b_\ell \ra }$ and  
{}From this it follows that 
\begin{equation}
e^{B_n}h^{\la A_\ell \ra }{}_{\la B_\ell }h_{B_{n-\ell }\ra }=e^{\la
A_\ell \ra }(+1)^{n-\ell }=O^{A_\ell }. 
\label{p-r004}
\end{equation}
It should also be noticed that from (\ref{Delta-ortho0-01}), 
\begin{equation}
h^{\la A_\ell \ra }{}_{\la A_\ell \ra }=(2\ell +1)\,, \label{pstfh-contract}
\end{equation}
can be also shown algebraically \cite{dGvLvW}\footnote{%
Note the contrast with (\ref{contr}).}.

Using these relations we obtain the inversion of the harmonic expansion 
(\ref{3.11}): 
\begin{eqnarray}
\tau (x^i,e^a)=\sum_{\ell =0}^\infty \tau _{A_\ell }(x^i)\,O^{A_\ell
}~\Leftrightarrow ~\tau _{A_\ell }(x^i)=\Delta _\ell ^{-1}\int_{4\pi }d\Omega 
\,O_{A_\ell }\,\tau(x^i,e^a)\,.  
\end{eqnarray}

\subsubsection{Addition of $O^{A_{\ell}}$}

The addition theorem for $O_{A_\ell}$ can be found from 
\begin{eqnarray}
O^{A_\ell }O^{\prime }{}_{A_\ell }=\sum_{k=0}^{[\ell /2]}\sum_{k^{\prime
}{}=0}^{[\ell /2]}B_{\ell k}B_{\ell k^{\prime }{}}h^{(A_{2k}}h_{(A_{2k^{\prime
}{}}}e^{A_\ell )}e^{\prime }{}_{A_\ell )}.  \label{add1} 
\end{eqnarray}
The resulting polynomial 
\begin{eqnarray}
L _\ell (X)\equiv O^{A_\ell }O^{\prime }{}_{A_\ell }=\sum_{m=0}^{[\ell
/2]}B_{\ell m}X^{\ell -2m}, 
\label{add01a}
\end{eqnarray}
is the natural polynomial that arises in the PSTF tensor approach.

\subsubsection{Double Integrals}

First, note that
\begin{eqnarray}
O^{\prime }{}_{A_\ell }e^{\prime }{}_{B_n}\int d\Omega O^{A_\ell
}e^{B_n} &=&\int d\Omega O^{\prime }{}_{A_\ell }O^{A_\ell }e^{\prime
}{}_{B_n}e^{B_n} \nonumber \\ 
&=&\beta _\ell {\frac{4\pi }{{n+\ell +1}}}{\frac{n!(n-\ell
+1)!!}{(n-\ell +1)!(n+\ell -1)!!}} 
\end{eqnarray}
on integrating a Legendre polynomial, where $n\geq l$ and $(n-\ell )$ is even,
so we can write $n-\ell =2m$ for $m$ an integer \footnote{This can
also be seen from using $F_{A_\ell } h^{(A_\ell  B_n)} = (n! / (n-\ell
)!!(n+\ell -1)!!) F_{A_\ell } h^{A_\ell  (B_\ell } h^{B_{n-\ell })}$,
the definition of $O^{A_\ell }$ in terms of $e^{A_\ell }$ and
evaluating $\int d \Omega F_{A_\ell } O^{A_\ell } e^{B_n}$ for $F_{A_\ell }$
PSTF.}. Consequently, on remembering  
\begin{eqnarray}
h^{\la A_\ell \ra }{}_{(B_n}h_{B_{n-\ell })}O^{\prime }{}_{A_\ell }e^{\prime
}{}^{B_n}=O^{\prime }{}_{A_\ell }e^{\prime }{}^{A_\ell }=\beta _\ell
,~~(e^{\prime}{}^ae^{\prime }{}_a)^{(n-\ell )}=+1\,,  
\end{eqnarray}
we find 
\begin{eqnarray}
\int d\Omega O^{A_\ell }e_{B_n}=4\pi \delta _n^{\ell
+2m}{\frac{n!(n-\ell +1)!!}{(n-\ell +1)!(n+\ell +1)!!}}h^{\la A_\ell
\ra }{}_{(B_\ell}h_{B_{n-\ell })}.  \label{dint}  
\end{eqnarray}
Here $m$ are positive integers. Also we will need 
\begin{eqnarray}
\int d\Omega _kO_{A_\ell }^{(k)}\int d\Omega _{k^{\prime }{}}O_{(k^{\prime
}{})}^{B_m} &=&\int d\Omega _k\int d\Omega _{k^{\prime
}{}}O_{A_\ell }^{(k^{\prime }{})}O_{(k)}^{B_m}=4\pi \int d\Omega
_kO_{A_\ell }^{(k)}O_{(k^{\prime }{})}^{B_m}  \nonumber \\
&=&4\pi \delta _\ell ^m\Delta _\ell h_{\la A_\ell\ra }{}^{\la B_\ell
\ra }\delta (e_k^a-e_{k^{\prime}{}}^a).  \label{ddint} 
\end{eqnarray}

\section{Mode relations}

\subsection{The curvature modified Helmholtz equation and the mode
recursion relation}

By successively applying the background 3-space Ricci identity, 
\begin{equation}
\D_{abA_\ell }Q-\D_{baA_\ell }Q=+\sum_{n=1}^\ell {\frac K{a^2}}\left( {\delta
_b^{b_n}h_{a_na}-\delta _a^{b_n}h_{a_nb}}\right) \D_{\bar{A_\ell }}Q,
\end{equation}
where $\bar{A_\ell }=a_1...a_{n-1}b_na_{n+1}...a_\ell $, {\it {\ i.e. }}, the
sequence of $\ell $ indices with the $n$-th one replaced with a
contraction, the following useful relations are found: 
\begin{eqnarray}
e^{(a_1}O^{A_\ell )}\D_{A_{\ell +1}}Q &=&e^{a_1}O^{A_\ell }\D_{A_\ell
}Q-\frac 13{\frac K{a^2}} {\frac{\ell ^2(\ell -1)}{(2\ell
-1)}}O^{A_{\ell -1}}\D_{A_{\ell -1}}Q, \label{rel-001} \\ 
h^{(a_1a_2}O^{A_{\ell -1})}\D_{A_{\ell +1}}Q &=&\left( {-k_{phys}
^2+\frac 13{\frac K{a^2}}{\frac{(\ell +2)}{(\ell -1)}}}\right)
O^{A_{\ell -1}}\D_{A_{\ell -1}}Q, \label{rel-002} \\
O^{A_\ell }\D_{cA_\ell }^{~~~c}Q &=&\left( {-k_{phys} ^2+\frac
12{\frac K{a^2}}\ell (\ell +3)}\right) O^{A_\ell }\D_{A_\ell }Q
\label{rel-003} \\ 
O^{A_\ell }\D_{cA_\ell }Q &=&O^{A_\ell }\D_{A_\ell c}Q+\frac 12{\frac
K{a^2}} \ell (\ell -1)O_c^{~A_{\ell -1}}\D_{A_{\ell -1}}Q.  \label{rel-004} 
\end{eqnarray}
Now consider $O^{A_\ell }\D_{~cA_\ell }^cQ$ from (\ref{rel-004}) we find 
\begin{equation}
O^{A_\ell }\D_{~cA_\ell }^cQ=O^{A_\ell }\D_{~A_\ell c}^cQ+\frac
12{\frac K{a^2}} \ell (\ell -1)O_c^{~A_{\ell -1}}\D_{~A_{\ell -1}}^cQ
\label{rel-005} 
\end{equation}
where the first term on the left of the equality above,
(\ref{rel-005}), can be reduced to one in
terms of $O^{A_\ell }\D_{A_\ell }Q$ using (\ref{rel-003}) while the
last term can also be rewritten in terms of $O^{A_\ell }\D_{A_\ell }Q$. Now on
dropping the $O^{A_\ell }$ and making the identification of $Q_{A_\ell
}=(-k_{phys} )^{-\ell}\D_{\la A_\ell \ra }Q$ it is then shown that the
$Q_{A_\ell }$ satisfy the {\it curvature-modified Helmholtz equation} 
\begin{equation}
\D^a\D_aQ_{\la A_\ell \ra }=(-k_{phys} ^2+{\frac K{a^2}}\ell (\ell
+2))Q_{\la A_\ell \ra}, \label{Acmhe}
\end{equation}
\ie, the Helmholtz equation with modified wavelength using as before 
$k_{phys} =k/a$ 
\begin{equation}
- \tilde{k}_\ell ^2={\frac 1{a^2}}(K\ell (\ell +2)-k^2).
\end{equation}

%
%
On using (\ref{Acmhe}) and taking the PSTF part of the lower
indices, $\D^a \D_{\la a} Q_{A_{\ell} \ra}$, and using the PSTF
tensor relation (\ref{eO-rec}) 
 to find : 
\begin{equation}
\D^a \D_{\la a} Q_{A_\ell \ra}  = {(\ell+1) \over (2 \ell+1)} 
\left( {- {k^2 \over a^2}} \right)  \left[ {1 - {K\over k^2} \ell
(\ell+2)} \right] Q_{\la A_{\ell} \ra}.
\label{PSTF-cmhe} 
\end{equation}

On substituting the first two relations (\ref{rel-001}, \ref{rel-002})
into the recursion relation for the PSTF tensors (\ref{PSTF_rec}), 
we find 
\begin{eqnarray}
e^a\D_a[G_\ell [Q]]=+k_{phys} \left[ {{\frac{\ell ^2}{(2\ell +1)(2\ell
-1)}}\left( {1-{\frac K{k^2}}(\ell ^2-1)}\right) G_{\ell
-1}[Q]-G_{\ell +1}[Q]}\right] , \label{AGl-recursion}
\end{eqnarray}

\subsection{Legendre tensors and the PSTF tensors}

One can immediately make the connection between this formulation and the one
usually used in terms of Legendre tensors, and see that the Legendre tensors
used in Wilson \cite{W83} in the coordinate basis (indicated by late
romans) can be related to irreducible representation $O^{A_\ell }$ in terms of
its associated tetrad frame $E_i^a=\left\{ u^a,e_\mu ^\alpha \right\} $. The
direction vectors $e^a$ in the triad with components $e_\mu ^\alpha $ are
related to $\raisebox{.4ex}{$\gamma$}^i$, the direction cosines used in the
Wilson-Silk coordinate basis treatment : 
\begin{equation}
P_{(\ell )}^{i_1i_2...i_\ell }(\gamma ^i)=(\beta
_\ell )^{-1}e_{a_1}^{i_1}e_{a_2}^{i_2}...e_{a_\ell }^{i_\ell }O^{A_\ell }(e^a).
\label{len-PSTF}
\end{equation}
This connection to the Legendre polynomials can be seen by using the
relation between spherical harmonics and the PSTF
tensor along with the addition theorem for the PSTF tensors : 
\begin{eqnarray}
O^{A_\ell }={\cal Y}_{\ell m}^{A_\ell }Y^{\ell m}(\Omega
),~~\Leftrightarrow ~\beta_\ell P_\ell (e^ae^{\prime }{}_a)=O^{A_\ell
}O^{\prime }{}_{A_\ell }. 
\end{eqnarray}
Multiply (\ref{PSTF_rec}) by $\beta _{\ell +1}^{-1}$ and use
(\ref{len-PSTF}) and (\ref{rec}) to find the recursion relation for
the Wilson-Silk \cite{WS,W83} Legendre polynomials  
\begin{eqnarray}
P_{(\ell +1)}^{i_1...i_\ell }={\frac{(2\ell +1)}{(\ell +1)}}\gamma
^{(i}P_{(\ell )}^{i_2...i_{\ell +1})}-{\frac \ell {(\ell +1)}}\gamma
^{(i_1i_2}P_{(\ell -1)}^{i_3...i_{\ell +1})}. \label{len-rec}
\end{eqnarray}
It is seen that the recursion relations (\ref{PSTF_rec}) for the
irreducible representation $O^{A_\ell }$ can be reduced to that of the Legendre
polynomial. This links the CGI-PSTF approach to the usual GI-Legendre 
tensor approach. 

\section{Plane-waves, spherical-waves and mode functions}

\subsection{Plane-wave and mode function relations}

We consider only flat, $K = 0$, universes at present. Each set of harmonic
functions $Q_{(k)}(x^\alpha)$ satisfying (\ref{eigen}) which 
has associated with it $k_{phys}= k/a$, the physical wavenumber, a variation
vector field, $q^a$, and a direction $e^a$ ( $e^a e_a=1,~e^a u_a=0$ )
determined by \footnote{The vector $e^a$ defined here is in general
different from that associated with the angular harmonic expansion in
(\ref{3.11}). When ambiguity can arise, we explicitly put in the
$k$-dependence : $q^a_{(k)}$, to signify both this dependence and the
definition of $e^a$ from (\ref{q_a}) : thus strictly we should
write, for example, $\D_a Q_{(k)}= Q_{(k)} q^{(k)}_a = Q_{(k)} q^{(k)}
e^{(k)}_a$. We will suppress the $k$ when this causes no ambiguity.}
\begin{equation}
\D_a Q = Q q_a,~~q_a = q e_a,~~ q^2 = q^q q_a  \label{q_a}
\end{equation}
the first equality defining $q_a(x^i)$ (but not necessarily so as to factor
out Q) and the second splitting it into its magnitude and direction. It
follows that 
\begin{equation}
\D^a \D_a Q = Q q^2 + Q \D^a q_a = Q(q^2 + e^a \D_a q + q \D^a e_a)
\label{q-eigen}
\end{equation}
so that (\ref{eigen}) becomes 
\begin{equation}
q^2 + e^a \D_a q + q \D^a e_a = - {k_{phys}^2}, ~~\iff \D^a q_a = - q^2 -
k_{phys}^2.  \label{99}
\end{equation}

Using the $K=0$ plane-wave eigenfunctions with associated
direction vector $e_a^{(k)}$: 
\begin{eqnarray}
Q(x^i,e_{(k)}^a)|_{flat}=\exp \left\{ {-ik_{phys} e_a^{(k)}x^a}\right\} ,
\label{plane-wave}
\end{eqnarray}
where $k_{phys} (k,t)=k/a$, expresses the temperature anisotropy (\ref
{3.10}) in terms of its plane-wave spatial Fourier Transform (\ref{t-pw}).
In this case 
\begin{eqnarray}
q_a=-ik_{phys} e_a^{(k)},~~\D_aq_b=0=\D_ak_{phys}
=\D_ae_b^{(k)},~~q^2=-{\frac{k^2}{a^2}}=-k_{phys} ^2 \label{q-defs}
\end{eqnarray}
holds in equation (\ref{q_a}) and (\ref{q-eigen}) respectively we find
(from (\ref{plane-wave})) 
\begin{equation}
\D_{\la A_\ell \ra }Q(x^i,e_{(k)}^a)|_{flat}=(-ik_{phys}
)^\ell O_{A_\ell }^{(k)}Q(x^i,e_{(k)}^a)|_{flat}
\end{equation}
where the $O_{A_\ell }^{(k)}$ are the PSTF tensors
associated with the direction $e_{(k)}^a$ in the tangent spaces on the
spatial section. Thus from (\ref{def1}) 
\begin{equation}
(Q(x^i,e_{(k)}^a)|_{flat})_{\la A_\ell \ra 
}=(-1)^\ell O_{A_\ell }^{(k)}Q(x^i,e_{(k)}^a)|_{flat} \label{102} 
\end{equation}
and (\ref{mode_exp}) to find (\ref{tAl-pw}).

\subsection{Radial expansion and mode function relations}

\begin{equation}
\D_a Q_\ell  = \D_a (R_{A_\ell } O^{A_\ell }) = O^{A_\ell } \D_a
R_{A_\ell } + R_{A_\ell } \D_a O^{A_\ell }, \label{DQ-l}
\end{equation}
which implies that 
\begin{equation}
\D_a \D^a Q_\ell  = (\D^a \D_a R_{A_\ell }) O^{A_\ell } + 2 (\D_a
R_{A_\ell } \D^a O^{A_\ell }) + R_{A_\ell } (\D_a \D_a O^{A_\ell }).
\label{D2Q-l}
\end{equation}
Now we need to work out $(\D_a R_{A_\ell })$, $\D^a \D_a R_{A_\ell }$,
$\D_a O^{A_\ell }$ and $\D_a \D^a O^{A_\ell }$, say (a), (b), (c), (d)
respectively. 
Calculating (a) : 
\begin{eqnarray}
\D_a R_{A_\ell }(r) = {\frac{\p  R_{A_\ell } }{\p  r}} \D_a r =
{\frac{\p  R_{A_\ell } }{\p  r}} e_a, \label{DRAl}
\end{eqnarray}
hence (b) follows: 
\begin{eqnarray}
\D^a \D_a R_{A_\ell }(r) = {\frac{\p ^2 R_{A_\ell } }{\p  r^2}} + \frac{2}{r}
{\frac{\p  R_{A_\ell } }{\p  r}} \,. \label{D2RAl}
\end{eqnarray}
Next, (c) is: 
\begin{eqnarray}
\D_a O^{A_\ell } = \frac{\ell }{r} p_a^{~\la a_\ell } O^{A_{\ell
-1}\ra } ~\Rightarrow e^a \D_a O^{A_\ell } = 0 \label{DOAl}
\end{eqnarray}
which gives (d) : 
\begin{eqnarray}
\D^a \D_a O^{A_\ell } &=& \frac{\ell }{r} (\D^a p_a^{~\la a_\ell })
O^{A_{\ell -1}\ra } + \frac{\ell }{r} p^{a \la a_\ell }(\D_a
O^{A_{\ell -1}\ra }) - \frac{\ell }{r^2} \D^a(r) p_a^{~\la a_\ell }
O^{A_{\ell -1}\ra }, 
\nonumber \\
&=& \frac{-2 \ell }{r^2} e^{\la a_\ell } O^{A_{\ell -1}\ra } +
\frac{\ell (\ell -1)}{r^2} p^{\la a_\ell a_{\ell -1}} O^{A_{\ell
-2}\ra } - \frac{\ell }{r^2} e^a p_a^{~\la a_\ell } O^{A_{\ell -1}\ra
},  \nonumber \\ &=& - \frac{\ell (\ell +1)}{r^2} O^{A_\ell }. \label{D2OAl}
\end{eqnarray}
Now put these in (\ref{D2Q-l}) to find, 
\begin{eqnarray}
\D_a \D^a Q_\ell  &=& \left[ {\ {\frac{\p ^2 R_{A_\ell } }{\p  r^2}} +
\frac{2}{r} {\frac{\p  R_{A_\ell } }{\p  r}}} \right] O^{A_\ell } + 2
\left[ {{\frac{\p  R_{A_\ell } }{\p  r}} e_a} \right] \left[
{\frac{\ell }{r} p_a^{~\la a_\ell } O^{A_{\ell -1}\ra }} \right]  \nonumber \\
&+& R_{A_\ell } \left[ {- \frac{\ell (\ell +1)}{r^2} O^{A_\ell }}
\right] = - k_{phys}^2 R_{A_\ell} O^{A_\ell}, \label{D2Ql-R}
\end{eqnarray}
which simplifies to 
\begin{eqnarray}
\frac{1}{r^2} {\frac{\p  }{\p  r}} \left( {r^2 {\frac{\p R_{A_\ell }
}{\p  r}}} \right) O^{A_\ell } + R_{A_\ell } \left[ {- \frac{\ell
(\ell +1)}{r^2} O^{A_\ell }} \right] = - k_{phys}^2 R_{A_\ell }
O^{A_\ell }.  \label{proto-bessel} 
\end{eqnarray}

Now we need to find $O^{A_\ell}_{(\chi)} D_{\la A_{\ell} \ra} Q$ in 
terms of $R_0$, then relate the resulting rodrigues formulae of
$R_\ell$ to get $R_\ell$ in terms of $R_0$ and hence $Q^{A_\ell}$ in
terms of $R_\ell$. In this regard, now in (\ref{103}), $\D_a e_b^{(k)}
= 0$ (see (\ref{99})) which implies 
$\D_a O_{A_\ell }^{(k)} =0$. Thus
\begin{equation}
\D_a R_{A_\ell }(r) = \sum_{k^a} {\frac{\p  R_\ell (k,r) }{\p  r}} e_a
O_{A_\ell }^{(k)} = {\frac{\p  R_0(k,r) }{\p  r}} e_a.  \label{104a}
\end{equation}
We can find $Q_{\la A_\ell \ra }$ from (\ref{def1}) 
obtaining 
\begin{eqnarray}
\D_{\la A_\ell \ra } Q &=& \sum_{m,k^{\nu}} \D_{\la A_\ell \ra }
R_{B_\ell } O^{B_\ell }_{(\chi)} = \sum_{m,k^{\nu}} \D_{\la A_\ell \ra
} R_m(r,k) O_{B_m}^{(k)} O^{(B_m)}_{(\chi)} = \D_{\la A_\ell \ra }
R_0(k,r),  \nonumber \\ &=& \D_{\la A_{\ell -1}} {\frac{\p  R_0 }{\p
r}} e_{a_\ell \ra } = \D_{\la A_{\ell -2}} \left( {\ {\frac{\p ^2 R_0
}{\p  r^2}} e_{a_{\ell -1}}e_{a_\ell \ra } + \frac{1}{r} p_{a_{\ell
-1} a_\ell \ra } {\frac{\p  R_0 }{\p  r}} }\right), \nonumber \\ &=&
\D_{\la A_{\ell -3}} \left( {{\frac{\p ^3 R_0 }{\p r^3}} e_{a_{\ell
-2}} e_{a_{\ell -1}} e_{a_\ell  \ra } + \frac{3}{r} {\frac{\p ^2 R_0
}{\p  r^2}} p_{a_{\ell -2} a_{\ell -1}} e_{a_{\ell }\ra } }\right.
\nonumber \\ &~&\left. {\ - {\frac{2 }{r}} {\frac{\p R_0 }{\p  r}}
p_{a_{\ell -2} a_{\ell -1}} e_{a_{\ell }\ra } - {\frac{1 }{r}}
{\frac{\p  R_0 }{\p  r}} (\D_{a_{\ell -2}}r ) p_{a_{\ell -1} a_{\ell
}\ra } }\right).  \label{r-m001}  
\end{eqnarray}
Now we note that 
\begin{eqnarray}
O^{A_\ell }_{(\chi)} \D_{\la A_\ell \ra } Q &=& O^{A_\ell }_{(\chi)}
\D_{\la A_\ell \ra }R_0(k,r) = O^{A_\ell }_{(\chi)} \D_{\la A_{\ell -2}}
\left( {{\frac{\p ^2 R_0}{\p  r^2}} - \frac{1}{r}
{\frac{\p  R_0 }{\p  r}}} \right) e_{a_{\ell -1}} e_{a_\ell
\ra }  \nonumber \\ &=& O^{A_\ell }_{(\chi)} (-r)^\ell \left( {\ -
{\frac{1 }{r}} {\frac{\p }{\p  r}} } \right)^\ell  R_0
e_{\la A_\ell \ra } \label{r-m002}  
\end{eqnarray}
which follows from $p_{ab} e^{\prime}{}^{\la a} e^{\prime}{}^{b\ra } = h_{ab}
O^{\prime}{}^{ab} - e_a e_b O^{\prime}{}^{ab} = - O_{ab} O^{\prime}{}^{ab}$ 
%
%
and that
\begin{equation}
R_\ell  = {\frac{r^\ell }{k_{phys}^\ell }} \left( {- \frac{1}{r}
{\frac{\p  }{\p  r}}} \right)^\ell  R_0\,. \label{Rl-rod}
\end{equation}
Used in (\ref{r-m002}) this gives that 
\begin{equation}
O^{A_{\ell}}_{(\chi)} \D_{\la A_{\ell} \ra} Q = (-k_{phys})^{\ell}
O^{A_{\ell}} O_{A_\ell}^{(k)} R_{\ell}. \label{QD-Rl} 
\end{equation}

\section{Mode mean square relations}

\subsection{Flat relations}

\begin{eqnarray}
\tau_{A_\ell } &=& (+i)^\ell  \int {k^2 d k \over (2 \pi)^3} 
\int d \Omega_k \tau_{\ell }(k,t) O^{(k)}_{A_\ell } 
\sum_{n=0}^{\infty} (+i)^n j_n(\lambda r) O_{(k)}^{B_n}
O_{B_n}^{(\chi)} \beta_n^{-1} (2n+1),  \nonumber \\
&=&  {(-1)^\ell \over 2 \pi^2} O^{(\chi)}_{A_\ell } 
\int k^2 dk \tau_\ell (k,t) j_\ell  (k \chi). \label{tau_Al_planewv}
\end{eqnarray}
Equivalently from 
\begin{equation}
\tau(x^i,e^a) = \sum_m \tau_m Q_m, ~~~Q_m = \sum_{k^a} R_{C_m D_m}
O^{D_m}_{(k)} O^{C_m}_{(\chi)}\,,
~~\mbox{and}~~
R_{C_m D_m} = R_\ell (k,r) h_{C_m D_m}.  \label{Rl_defn}
\end{equation}
Invert the multipole expansion 
\begin{equation}
\tau_{A_\ell } = \Delta_\ell ^{-1} \int d \Omega_k O_{A_\ell }^{(k)} \left\{
\sum_{m,k^a} \tau_m R_{C_m B_m} O^{B_m}_{(k)} O^{C_m}_{(\chi)} \right\},
\end{equation}
to find, on using (\ref{ddint}) , that 
\begin{eqnarray}
\tau_{A_\ell } &=& {1 \over 2 \pi^2} \Delta_\ell ^{-1} \sum_m \int k^{\prime}{}^2
dk^{\prime}{}\tau_m(k^{\prime}{},t) R_{C_m B_m} O^{C_m}_{(\chi)}
\left[ {\delta^m_\ell  \Delta_\ell  h_{\la A_\ell \ra }^{~~\la B_\ell
\ra }} \right], 
\end{eqnarray}
and hence that 
\begin{eqnarray}
\tau_{A_\ell } = {1 \over 2 \pi^2} \int k^{\prime}{}^2 dk^{\prime}{}\tau_\ell
(k^{\prime}{},t) R_{A_\ell  C_\ell } O^{C_\ell }_{(\chi)}\,.
\end{eqnarray}
Using (\ref{Rl_defn}) this becomes 
\begin{eqnarray}
\tau_{A_\ell } = {1 \over 2 \pi^2} O_{A_\ell }^{(\chi)} \int k^{\prime}{}^2
dk^{\prime}{}\tau_\ell (k^{\prime}{},t) R_\ell (k^{\prime}{},\chi),
\label{Atau_Al_radial}
\end{eqnarray}

\subsection{Constant curvature relations}

We now have that 
\begin{eqnarray}
\int d \Omega_{(\chi)} O^{A_\ell }_{(\chi)} G_\ell [Q] &=& \delta^\ell
_m \Delta_\ell Q^{A_\ell }, \label{OGl} \\ \int d \Omega_{(\chi)} G_\ell [Q]
G_m[Q] &=& \delta_\ell ^m Q^{A_\ell } Q_{A_\ell } \Delta_\ell , \label{GlGm} \\
\int d \Omega_{(k)} G_\ell [Q] G^{\prime}{}_m[Q] &=& \int d
\Omega_{(k)} O^{A_\ell }_{(\chi)} R_\ell O_{A_\ell }^{(k)}
O^{B_m}_{(\chi)} R_m O_{B_m}^{(k)} = \delta_m^\ell \Delta_\ell  R_\ell
^2 O_{A_\ell }^{(\chi)} O^{\prime}{}^{A_\ell }_{(\chi)}  \nonumber \\
~~~~~&\Rightarrow& \int d \Omega_{(k)} G_\ell G_m = \delta^{\ell}_{m}
\Delta_\ell ^2  {\frac{(2\ell +1) }{4 \pi}} R_\ell ^2.  \label{Gl-norm1} 
\end{eqnarray}

Furthermore, we have that from the recursion relations (\ref {Gl-recursion}) 
\begin{equation}
e^a_{(\chi)} \D_a [G_\ell ] = + k_{phys} [ \alpha_\ell ^2 G_{\ell -1}
- G_{\ell +1}] 
\end{equation}
where
\begin{equation}
(\alpha_\ell )^2 = {\frac{\ell ^2 }{(2\ell +1)(2\ell -1)}} \left({\ 1
- {\frac{K }{k^2}} (\ell ^2-1)}\right)
\end{equation}
and using \cite{WS} {\footnote{The idea is to use this to fix the
normalization of $\int d \Omega_{(k)} G_\ell [Q] = D_\ell  \int d
\Omega_{(k)} Q(x^i,e^a_{(k)}) O^{A_\ell }_{(k)} O_{A_\ell }^{(\chi)}$.}}, 
\begin{equation}
\int d \Omega_k \int d x^{\nu} e^a_{(\chi)} \D_a [G_\ell  G_{\ell -1}] = 0
\end{equation}
to find
\begin{equation}
\int d \Omega^{(k)} \left< | G_n |^2 \right> = (\alpha_n)^2 \int d
\Omega_{(k)} \left< |G_{n-1}|^2 \right>. \label{rec_norm_Gn}
\end{equation}
Here we have defined the mean square to pick out the power spectrum which
is a function only of the absolute value of the wavelength for a Gaussian
distribution (there is no directional dependence, the modulus is only
dependent on the wave number).

\section{1+3 Ortho-Normal Tetrad relations}

We use an orthonormal tetrad approach (cf. \cite
{ETMa}). Consider an orthonormal tetrad basis $E_a$ with components $%
E_a^i(x^j)$ relative to a coordinate basis; here indices a,b,c... , that is
early letters, are used for the tetrad basis, while late letters i,j,k,...
are used for the coordinate basis. The differential operators $\partial_a =
E^i_a \partial_i$ are defined by the inverse basis components 
\begin{equation}
E^a_i E^j_a = \delta ^j_i~~ \Leftrightarrow ~~E^a_i E^i_b = \delta^a_b\,.
\end{equation}
The tetrad components of a vector $X^i$ are ${X}^a = {E}^a_{i} X^i$, and
similarly for any tensor. Tetrad indices are raised and lowered using the
tetrad components of the metric 
\begin{equation}
g_{ab}=g_{ij} E^i_a E^j_b = \mbox{diag}(-1,+1,+1,+1)\,, ~~g^{ab}g_{bc}=
\delta^a_c\,  \label{tdef}
\end{equation}
the form of these components being the necessary and sufficient condition that
 the tetrad basis vectors used are orthonormal, which we will always assume.

For an observer with 4-velocity $u^a$, there is a preferred family of
orthonormal tetrads associated with $u^a$ {\it {\ i.e. }} a frame for which
the time-like tetrad basis ${E}_0$ is parallel to the velocity $u^a$. In
such a tetrad basis 
\begin{eqnarray}
u^a &=& \delta^a_0 \\
h_{ab} &=& \mbox{diag}(0,+1,+1,+1)
\end{eqnarray}
All our work is based on such a tetrad, which leads to a preferred set of
associated rotation coefficients. In paper 1, the form of these rotation
coefficients is unimportant, so we defer their consideration to Paper 2.
The issue for the present is that we have a preferred family of local
orthonormal frames at each point (usually matter flow aligned), and carry
out our algebraic analysis of observational quantities relative to that
orthonormal frame.

\[ \]

\end{document}